\documentstyle[psfig,12pt]{article}
\def\TL{\hfil$\displaystyle{##}$}
\def\TR{$\displaystyle{{}##}$\hfil}

\def\comment#1{}
\def\fixit#1{}

\def\mop#1{\mathop{\rm #1}\nolimits}

\def\vol{\mop{vol}}

\def\tr{\mop{tr}}

\def\overleftrightarrow#1{\vbox{\ialign{##\crcr
     $\leftrightarrow$\crcr\noalign{\kern-0pt\nointerlineskip}
     $\hfil\displaystyle{#1}\hfil$\crcr}}}

\def\lsim{\mathrel{\mathstrut\smash{\ooalign{\raise2.5pt\hbox{$<$}\cr\lower2.5pt\hbox{$\sim$}}}}}
\def\gsim{\mathrel{\mathstrut\smash{\ooalign{\raise2.5pt\hbox{$>$}\cr\lower2.5pt\hbox{$\sim$}}}}}

\def\sqr#1#2{{\vcenter{\vbox{\hrule height.#2pt
         \hbox{\vrule width.#2pt height#1pt \kern#1pt
            \vrule width.#2pt}
         \hrule height.#2pt}}}}
\def\square{\mathop{\mathchoice\sqr56\sqr56\sqr{3.75}4\sqr34\,}\nolimits}

\def\href#1#2{#2}  

\def\lbldef#1#2{\expandafter\gdef\csname #1\endcsname {#2}}
\def\eqn#1#2{\lbldef{#1}{(\ref{#1})}%
\begin{equation} #2 \label{#1} \end{equation}}
\def\eqalign#1{\vcenter{\openup1\jot
    \halign{\strut\span\TL & \span\TR\cr #1 \cr
   }}}
\def\eno#1{(\ref{#1})}

\def\comment#1{  \begin{raggedright}{\tt [#1]}\end{raggedright}}
\begin{document}
\baselineskip=15.5pt
\pagestyle{plain}
\setcounter{page}{1}
\renewcommand{\theequation}{\thesection.\arabic{equation}}
\begin{titlepage}
\bigskip
\rightline{}
\rightline{CALT-68-2328}
\rightline{CITUSC/01-014}
\rightline{NSF-ITP-01-31}
\rightline{PUPT-1984}
\rightline{hep-th/0105047}
\bigskip\bigskip\bigskip\bigskip
\centerline{\Large \bf {Stability of $AdS_p \times M_q$ Compactifications}} 
\medskip
\centerline{\Large \bf {Without Supersymmetry}}
\bigskip\bigskip
\bigskip\bigskip

\centerline{\large Oliver DeWolfe${}^1$, Daniel Z. Freedman${}^{2,3}$,
Steven S. Gubser${}^{4,5}$, }
\smallskip
\centerline{\large Gary T. Horowitz${}^{1,6}$ and Indrajit Mitra${}^5$}
\bigskip
\centerline{\em ${}^1$Institute for Theoretical Physics, UCSB, Santa Barbara,
CA 93106}
\medskip
\centerline{\em ${}^2$Caltech-USC Center for Theoretical Physics, USC, Los Angeles, CA 90089}
\medskip
\centerline{\em ${}^3$Department of Mathematics and Center for Theoretical Physics}
\centerline{\em Massachusetts Institute of Technology, Cambridge, MA 02139}
\medskip
\centerline{\em ${}^4$Lauritsen Laboratory of Physics, Caltech, Pasadena, CA 91125}
\medskip
\centerline{\em ${}^5$Joseph Henry Laboratories, Princeton University, Princeton, NJ 08544}
\medskip
\centerline{\em ${}^6$Department of Physics, UCSB, 
Santa Barbara, CA 93106}
\medskip
\bigskip\bigskip


\begin{abstract}
We study the stability of Freund-Rubin compactifications, $AdS_p
\times M_q$, of $p+q$-dimen\-sional gravity theories with a $q$-form
field strength and no cosmological term.  We show that the general
$AdS_p \times S^q$ vacuum is classically stable against small
fluctuations, in the sense that all modes satisfy the
Breitenlohner-Freedman bound.  In particular, the compactifications
used in the recent discussion of the proposed bosonic M-theory are
perturbatively stable.  Our analysis treats all modes arising from the
graviton and the $q$-form, and is completely independent of
supersymmetry.  From the masses of the linearized perturbations, we
obtain the dimensions of some operators in possible holographic dual
CFT's.  Solutions with more general compact Einstein spaces need not
be stable, and in particular $AdS_p\times S^n\times S^{q-n}$ is
unstable for $q<9$ but is stable for $q\ge 9$.  We also study the
$AdS_4\times S^6$ compactification of massive type IIA supergravity,
which differs from the usual Freund-Rubin compactification in that
there is a cosmological term already in ten dimensions.  This
nonsupersymmetric vacuum is unstable.

\medskip
\noindent
\end{abstract}
\end{titlepage}


\section{Introduction}

The discovery of the $AdS$/CFT correspondence
\cite{juanAdS,gkPol,witHolOne} (for a review see \cite{MAGOO}) has led
to renewed interest in the stability of geometries of the form $AdS_p
\times M_q$ where $AdS_p$ is anti-de Sitter spacetime and $M_q$ is an
Einstein space with positive Ricci tensor.  Solutions of this type
with a $q$-form field strength on $M_q$ were first considered in
higher dimensional supergravity theories by Freund and Rubin
\cite{FreundRubin}.  Due to the negative curvature of $AdS$,
perturbative stability does not require the absence of all tachyonic
modes.  Instead, as Breitenlohner and Freedman (BF) first showed,
scalars with $m^2 < 0$ may appear as long as their masses do not fall
below a bound set by the curvature scale of $AdS$ \cite{BF}.  The
issue of stability is important for understanding a possible dual
conformal field theory (CFT) description.  For stable solutions, the
spectrum of masses directly yields the dimensions of certain operators
in such a CFT.  Unstable solutions can still have a dual CFT
description \cite{Eva}, but the physics is clearly very different.

It is well known that for the standard ten and eleven dimensional
maximally supersymmetric supergravity theories, 11D SUGRA on $AdS_4
\times S^7$ or $AdS_7 \times S^4$ and Type IIB SUGRA on $AdS_5 \times
S^5$ are all stable. However, these solutions are all supersymmetric
(SUSY), and simple nonsupersymmetric vacua like $AdS_4 \times M_n
\times M_{7-n}$ \cite{DNP} and $AdS_7\times S^2\times S^2$
\cite{BerkoozRey} are known to be unstable.  Furthermore, the SUSY
examples have modes which either saturate the BF bound, or are very
close to saturating it.  This raises the question of the role that
SUSY plays in ensuring stability of vacua of this type.  (For earlier
discussions of this question see {\em e.g.}~\cite{DNP},
\cite{BerkoozRey}, \cite{EvaShamit}.)  This issue is of particular
interest in light of the recent proposal of bosonic M-theory
\cite{HS}, a 27-dimensional theory which was hypothesized to appear as
the strong-coupling limit of the bosonic string. Its low energy limit
is assumed to be gravity coupled to a four-form field strength, which
admits solutions of the form $AdS_4 \times S^{23}$ and $AdS_{23}
\times S^4$. It was suggested that with these boundary conditions,
bosonic M-theory might be holographically described by a (2+1)- or
(21+1)-dimensional CFT. Thus, it is important to determine whether
these solutions are stable.

One argument for the stability of $AdS_4 \times S^{23}$ \cite{HS} and
more generally $AdS_p \times S^q$ is that these backgrounds are the
near-horizon geometries of extremal black branes. However this is not
completely satisfying for two reasons. First, although we expect
extremal black branes to be stable, the appropriate positive mass
theorem (stating roughly $M\ge Q$) has never been
proven.\footnote{Interestingly enough, if one tries to adapt Witten's
spinorial approach, one succeeds only in the SUSY cases \cite{GHT}.}
Second, as we will discuss later, one can construct extremal black
brane solutions with unstable near horizon geometry by placing branes
at the apex of appropriate cones.  So, one needs to examine stability
directly.

In this paper, we study the stability of general solutions of the form
$AdS_p\times M_q$ in a theory of gravity coupled to a $q$-form field
strength.  When one expands the field equations to linear order, there
are several types of modes. Some immediately decouple from the others,
while the rest mix and must be diagonalized.  {\em A priori}, since the
fundamental fields in $p+q$ dimensions are massless, and adding
dependence on $M_q$ should increase the mass, one might expect that
the modes that don't mix should always be stable. Masses violating the
BF bound might be expected, however, to arise in diagonalizing the
coupled fluctuations --- indeed, this is the origin of the modes that
saturate or come very near to saturating the BF bound in the SUSY
examples, so one might think that the absence of supersymmetry could
push them over the edge.

Surprisingly, this is not what we find. It turns out that for any $p$
and $q$ and any Einstein space $M_q$, the coupled modes are always
stable.  Moreover, for $S^q$ the lowest mass either saturates ($q$
odd) or almost saturates ($q$ even) the BF bound.  This is not to say,
however, that an arbitrary $AdS_p \times M_q$ background is stable.
The dangerous mode turns out to be an unmixed scalar coming from the
transverse, traceless metric perturbation on $M_q$.  This is the only
mode which is sensitive to the choice of Einstein manifold $M_q$.  If
$M_q$ is the round sphere $S^q$, it is easy to show that this mode is
stable. In particular, the spacetimes of interest for bosonic
M-theory, $AdS_4\times S^{23}$ and $AdS_{23}\times S^4$, are
stable. However, if $M_q= M_n\times M_{q-n}$ and $q<9$, there is a
mass violating the BF bound, corresponding to a mode which makes one
factor grow while the other shrinks.  This generalizes the
instabilities of $AdS_4 \times M_n \times M_{7-n}$ and $AdS_7\times
S^2\times S^2$, but also shows that this instability is limited to low
dimensions. For $q\ge 9$, $AdS_p\times S^n\times S^{q-n}$ can be shown
to be stable.  The significance of the critical dimension $q=9$ is not
clear; it is sufficiently large that stable products cannot be
realized in superstring/M-theory.

The massive type IIA supergravity has a nonsupersymmetric vacuum of
the form $AdS_4\times S^6$ \cite{RomansIIA}, whose stability, to our
knowledge, has never been investigated.  We also study this case and
show that the solution is unstable, with two modes violating the BF
bound. To our knowledge, this is the first example of a theory where
 the product of AdS and a round sphere
is unstable.  The analysis is more involved here since there is a dilaton
which mixes with some of the other modes, further complicating the
coupled sector.  Instabilities for more general $AdS_4 \times M_6$ can
arise in several ways, but we show in particular that they do occur
for $AdS_4 \times S^n \times S^{6-n}$.

There is a vast literature on Kaluza-Klein theories, much of it in the
context of higher dimensional supergravity, including a comprehensive
review \cite{Duffetal}. Our treatment of the harmonic analysis of
fluctuations about $AdS_p \times M_q$ is most closely modeled on
\cite{Castellanietal,Kimetal,vanN}, and we have also consulted
\cite{DNP} and \cite{Biranetal}.  In Section~\ref{Freund} we present
the general $AdS_p \times M_q$ background solution.  The harmonic
expansions for fluctuations and their linear equations of motion are
discussed in Section~\ref{Linearize}. The mass spectra of the various
fluctuations are analyzed in
Sections~\ref{ScalarSec}-\ref{UnstableSec}.  The more complicated case
of massive type IIA supergravity is discussed in
Section~\ref{MassiveIIA}. The $AdS_p$ mass spectra determine the
dimensions of operators in hypothetical $CFT_{p-1}$ dual field
theories, and this is discussed in Section~\ref{CFTDuals}. In
Section~\ref{Endpoint}, we show that for some of the
the unstable cases, the total
energy (in the full nonlinear theory) is unbounded from below. We also
speculate on the implications of our
results for the stability of certain extremal black brane solutions.
Conventions and properties of various differential operators are
collected in an appendix.

\section{Freund-Rubin Backgrounds}
\label{Freund}

We start by considering classical $D =p+q$ dimensional gravity
theory coupled to a $q$-form field strength.  The action is given by:
 \eqn{Action}{
 S = \int{d^{p}x d^{q}y \sqrt{-g} \left(R - {1 \over {2 q!}} F_{q}^2 \right)} \,,
  }
which leads to the equations of motion 
 \eqn{MetricEOM}{
R_{MN} = {1 \over {2 (q-1)!}} F_{M P_2 \cdots P_q} F^{\;\; P_2 \cdots P_q}_N - {{(q-1)} \over {2 (D-2) q!}} g_{MN}F_{q}^2 \,,
 }
 \eqn{FormEOM}{
 d * F_{q} =0 \,,
 }
This theory supports a Freund-Rubin solution with the product metric
 \eqn{ProdSpace}{
 ds^2 = ds_{AdS_p}^2 + ds_{M_q}^2 \,,
 } 
describing a product of $p$-dimensional anti-de Sitter space with an Einstein manifold:
 \eqn{BackgroundAdS}{
 R_{\mu \nu} = -{{(p-1)} \over L^2}g_{\mu \nu} \,,
 }
 \eqn{BackgroundM}{
 R_{\alpha \beta} = {{(q-1)} \over R^2}g_{\alpha \beta} \,, 
 }
and a background field strength on the compact space:
 \eqn{BackgroundF}{
 F_{q} = c \vol_{M_q} \,.
 }
We use $M, N, \ldots$ for indices on the full $D$-dimensional
spacetime, while $\mu, \nu, \ldots$ are indices on $AdS$ and $\alpha, \beta, \ldots$ are indices on $M_q$. The equations of motion
(\ref{MetricEOM}), (\ref{FormEOM}) relate the length scales
$L$ and $R$ and the constant $c$:
 \eqn{Valueofc}{
 c^2 = {{2(D-2)(q-1)} \over {(p-1) R^2}} \,,
}
 \eqn{RadiusRatio}{
 L = {{p-1} \over {q-1}} R \,.
 }
In the following six sections we shall study fluctuations of $g_{MN}$
and $F_q$ around this background.  Among other things, we will
conclude that the background is stable against these fluctuations when
$M_q = S^q$, for arbitrary $p>2$ and $q>1$.  If one wishes to embed
the action (\ref{Action}) in a larger theory with additional fields,
stability must be verified separately for the new modes.  However let
us note that the most tachyonic modes in the well-studied vacua of
ten- and eleven-dimensional supergravities generally come from
precisely the fields which support the solution.  Thus, when these
most ``dangerous'' modes come out stable, it suggests that the
background is probably stable against all fluctuations.

\setcounter{equation}{0}
\section{Linearized equations of motion}
\label{Linearize}

\subsection{Fluctuations}

We are interested in studying the stability of linearized fluctuations
around the background (\ref{ProdSpace}), (\ref{BackgroundF}).  As we
have discussed, anti-de Sitter space is stable even in the presence
of tachyonic scalar fields, as long as their masses do not violate the
Breitenlohner-Freedman bound:
 \eqn{Bound}{
 m^2 L^2 \geq - {{(p-1)^2} \over 4} \,.
 }
The possibility that some tachyons could be acceptable in $AdS_4$ was
first pointed out by Breitenlohner and Freedman \cite{BF}, and extended to
$AdS_p$ by \cite{ExtendBF}.  See also \cite{BFtwo,Townsend} for early
developments of this idea.

We consider the linearized fluctuations 
\begin{eqnarray}
 \delta g_{\mu \nu} &=& h_{\mu \nu} = H_{\mu \nu}  - {1 \over {p-2}} g_{\mu \nu} h^\alpha_\alpha \,, \label{MetricAdS} \\
\delta g_{\mu \alpha} &=& h_{\mu \alpha} \,, \quad \quad \delta
g_{\alpha \beta} = h_{\alpha \beta} \,, \quad \quad\delta A_{q-1} =
a_{q-1} \,, \quad\delta F_q \equiv f_q = da_{q-1} \,,
\end{eqnarray}
where we have defined a standard linearized Weyl shift on $h_{\mu
\nu}$ in (\ref{MetricAdS}), and $F_q = d A_{q-1}$.  It will be useful
to decompose $H_{\mu \nu}$ and $h_{\alpha \beta}$ into trace and
traceless parts:
\begin{eqnarray}
 H_{\mu \nu} = H_{(\mu \nu)} + {1 \over p} g_{\mu \nu} H^\rho_\rho \,,
 \quad \quad h_{\alpha \beta} = h_{(\alpha \beta)} + {1 \over q}
 g_{\alpha \beta} h^\gamma_\gamma \,,
\end{eqnarray}
where $g^{\mu \nu} H_{(\mu \nu)} = g^{\alpha \beta} h_{(\alpha \beta)}
= 0$.  To (mostly\footnote{ Besides unfixed $p$-dimensional
diffeomorphisms and gauge transformations, extra conformal
diffeomorphisms remain on $S^q$.  These are related to the elimination
of a $k=1$ mode in the coupled scalar sector, as in section
\ref{ScalarSec}; for a discussion, see \cite{Kimetal}.}) fix the
internal diffeomorphisms and gauge freedom, we impose the de
Donder-type gauge conditions
\begin{eqnarray}
\label{DDGauge}
\nabla^\alpha h_{(\alpha \beta)} = \nabla^\alpha h_{\alpha \mu} = 0 \,,
\end{eqnarray}
as well as the Lorentz-type conditions
\begin{eqnarray}
\label{LorentzGauge}
\nabla^\alpha a_{\alpha \beta_2 \ldots \beta_{q-1}} = \nabla^\alpha a_{\alpha \beta_2 \ldots \beta_{q-2} \mu} = \cdots = \nabla^\alpha a_{\alpha \mu_2 \ldots \mu_{q-1}} = 0  \,.
\end{eqnarray}
A generic gauge potential $a_{\alpha_1 \ldots \alpha_n \mu_{n+1}
\ldots \mu_{q-1}}$, viewed as an $n$-form on $M_q$ with additional
$AdS_p$ indices, can be expanded as the sum of an exact, a co-exact
and a harmonic form on $M_q$ by the Hodge decomposition theorem.  The Lorentz
conditions (\ref{LorentzGauge}), which state that the form is
co-exact, require the exact form in the decomposition to vanish, and hence the potentials can be expanded as co-exact forms (curls) and harmonic forms:
\begin{eqnarray}
\label{AsCurls}
a_{\beta_1 \ldots \beta_n \mu_{n+1} \mu_{q-1}} = \epsilon^{\alpha_1
\alpha_2 \ldots
\alpha_{q-n}}_{\;\;\;\;\;\;\;\;\;\;\;\;\;\;\;\;\;\;\;\; \beta_1 \ldots
\beta_n} \nabla_{\alpha_1} b_{\alpha_2 \ldots \alpha_{q-n} \mu_{n+1} \ldots
\mu_{q-1}} + \beta_{\beta_1 \ldots \beta_n \mu_{n+1} \mu_{q-1}}^{harm}\,.
\end{eqnarray}
When the compact space is an $S^q$ there are no nontrivial harmonic forms, but they can appear for other $M_q$.
In a compact
notation, we may write (\ref{LorentzGauge}) and (\ref{AsCurls}) as
\begin{eqnarray}
\label{CompactGauge}
d_q \! *_q \! a = 0 \rightarrow a = *_q d_q b + \beta^{harm}\,,
\end{eqnarray}
where $d_q$ and $*_q$ are the exterior derivative and Hodge dual with
respect to the $M_q$ space only.  

With these gauge choices, we may expand the fluctuations in spherical
harmonics as
\begin{eqnarray}
H_{(\mu \nu)}(x,y) &=& \sum_I H^I_{(\mu \nu)}(x) Y^I(y) \,, \quad \quad
H^\mu_\mu(x,y) = \sum_I H^I(x) Y^I(y) \,, \\
h_{(\alpha \beta)}(x,y) &=& \sum_I \phi^I(x) Y^I_{(\alpha \beta)}(y)
\,, \quad \quad h^\alpha_\alpha(x,y) = \sum_I \pi^I(x) Y^I(y) \,, \\
h_{\mu \alpha}(x,y) &=&  \sum_I B^I_\mu(x) Y^I_\alpha(y) \,, \\
a_{\beta_1 \ldots \beta_{q-1}} &=& \sum_I b^I(x) \; \epsilon^{\alpha}_{\;\;
\beta_1 \ldots \beta_{q-1}} \nabla_\alpha Y^I(y) \,, \label{GaugeFluct1}\\
a_{\mu \beta_2 \ldots \beta_{q-1}} &=&  \sum_I b^I_\mu(x) \; \epsilon^{\alpha \beta}_{\;\;\;\;
\beta_2 \ldots \beta_{q-1}} \nabla_{[\alpha} Y^I_{\beta]}(y) 
+ \sum_h \beta_\mu^h(x) \; \epsilon^{\alpha \beta}_{\;\;\;\;
\beta_2 \ldots \beta_{q-1}} Y^h_{[\alpha \beta]}
\,, \label{GaugeFluct2}\\
&\vdots& \nonumber \\
a_{\mu_1 \ldots \mu_{q-1}} &=& \sum_I b^I_{\mu_1 \ldots \mu_{q-1}}(x) Y^I(y) \,, \label{AllAdSb}
\end{eqnarray}
where $I$ in each case is a generic label running over the possible
spherical harmonics of the appropriate tensor type, and $h = 1 \ldots
b^n(M_q)$ runs over the harmonic $n$-forms on $M_q$ for the gauge
field with ($n-$1) $AdS_p$ indices.  We have not included a term
$\beta(x)$ in (\ref{GaugeFluct1}) since compact Riemannian Einstein
spaces with positive curvature cannot possess harmonic one-forms; this
is proved in the appendix.  We will also find it convenient to define
\begin{eqnarray}
b(x,y) \equiv \sum_I b^I(x) Y^I(y) \,, \quad \quad b_{\mu\alpha}(x,y)
\equiv \sum_I b^I_\mu(x) Y^I_\alpha(y) \,. 
\end{eqnarray}

\subsection{Einstein equations and coupled form equations}
\label{EquationSec}

We now consider the Einstein equations to linear order in
fluctuations, as well as the form equations that mix with the
graviton; the uncoupled form equations will be treated in
section~\ref{UncoupledSec}.  We use the following notation: $\square_x
\equiv g^{\mu\nu} \nabla_\mu \nabla_\nu$, $\square_y \equiv g^{\alpha
\beta} \nabla_\alpha \nabla_\beta$, and ${\rm Max}\ B_\mu \equiv
\square_x B_\mu - \nabla^\nu \nabla_\mu B_\nu$ is the Maxwell operator
acting on vectors on $AdS_q$.  Additionally, $\Delta_y \equiv -
(d_q^\dagger d_q + d_q d_q^\dagger)$ is the Laplacian\footnote{The
negative sign is standard in the Kaluza-Klein literature.} acting on
differential forms on $M_q$; for vectors, the explicit form is
$\Delta_y Y_\alpha \equiv \square_y Y_\alpha - R_\alpha^{\;\; \beta}
Y_\beta$.  Further, $f \cdot \epsilon \equiv f_{\alpha_1 \cdots
\alpha_q} \epsilon^{\alpha_1 \cdots \alpha_q} / q!$.

For convenience, we present the linearized Ricci tensor in our conventions:
 \eqn{RicciExpand}{\eqalign{
  R^{\;\; (1)}_{MN} &= -{1 \over 2}[(\square_x + \square_y)h_{MN} + \nabla_M \nabla_N h_{P}^{P} - \nabla_M \nabla^P h_{PN} - \nabla_N \nabla^P h_{PM} \cr
 &\qquad{}- 2 R_{MPQN} h^{PQ} - R_{M}^{\;\; P}h_{NP} - R_{N}^{\;\; P} h_{MP}] \,.
 }}
We employ Einstein's equations in their Ricci form, $R_{MN} = \bar{T}_{MN}$
with $\bar{T}_{MN} \equiv T_{MN} + {1 \over 2-D}\, g_{MN} \, T^P_P$.
For $R_{\mu \nu}$ we find
\begin{eqnarray}
\nonumber R^{\;\; (1)}_{\mu\nu} = -{1 \over 2}[(\square_x + \square_y)
(H_{\mu\nu} - {1 \over p-2} g_{\mu\nu} h^\gamma_\gamma) + \nabla_\mu
\nabla_\nu (H^\rho_\rho - {2 \over p-2} h^\gamma_\gamma) \\-
\nabla_\mu \nabla^\rho (H_{\rho\nu} - {1 \over p-2} g_{\rho\nu}
h^\gamma_\gamma) - \nabla_\nu \nabla^\rho (H_{\rho\mu} - {1 \over p-2}
g_{\rho\mu} h^\gamma_\gamma) - 2 R_{\mu\rho\sigma\nu} (H^{\rho\sigma}
- {1 \over p-2} g^{\rho\sigma} h^\gamma_\gamma) \\ - R_{\mu}^{\;\;
\rho} (H_{\rho\nu} - {1 \over p-2} g_{\rho\nu} h^\gamma_\gamma) -
R_{\nu}^{\;\; \rho} (H_{\rho\mu} - {1 \over p-2} g_{\rho\mu}
h^\gamma_\gamma)] \,, \nonumber
\end{eqnarray}
which must be equal to
\begin{eqnarray}
\bar{T}^{\;\; (1)}_{\mu\nu} = - {c^2 (q-1) \over 2 (D-2)} h_{\mu\nu} - {q (q-1)
c^2 \over 2 (D-2)q!} g_{\mu\nu} (- h^{\alpha \beta})
\epsilon_{\alpha \gamma_2 \cdots \gamma_q} \epsilon_{\beta}^{\;\;\;
\gamma_2 \cdots \gamma_q}  - {c (q-1) \over (D-2)} g_{\mu\nu}\
(f \cdot \epsilon) \,,
\end{eqnarray}
resulting in the equation
\begin{equation}\eqalign{
-{1 \over 2}[(\square_x + \square_y) H_{\mu\nu} + \nabla_\mu
\nabla_\nu H_\rho^\rho - \nabla_\mu \nabla^\rho H_{\rho\nu} -
\nabla_\nu \nabla^\rho H_{\rho\mu} - 2 R_{\mu\rho\sigma\nu}
H^{\rho\sigma} - R_{\mu}^{\;\; \rho} H_{\rho\nu} - R_{\nu}^{\;\; \rho}
H_{\rho\mu}] \cr{} + {1 \over 2(p-2)} g_{\mu\nu} (\square_x + \square_y)
h^\gamma_\gamma - {(q-1)^2 \over {(p-2) R^2}} g_{\mu\nu} h^\gamma_\gamma
+ {(q-1)^2 \over (p-1) R^2} H_{\mu\nu} + {q-1 \over D-2} g_{\mu\nu} \square_y cb = 0 \,. }
\end{equation}
For linearized $R_{\mu\alpha}$, we find
\begin{eqnarray}
R_{\mu\alpha}^{\;\; (1)} = -{1\over 2} [ \square_x h_{\mu\alpha} -
\nabla_\mu \nabla^\nu h_{\nu \alpha} - R_{\mu}^{\;\; \nu} h_{\nu
\alpha} + \square_y h_{\mu \alpha} - R_\alpha^{\;\; \beta} h_{\beta
\mu} \\ - \nabla_\alpha^\nu h_{\nu\mu} + \nabla_\mu \nabla_\alpha
(H^\rho_\rho - {2 \over p-2} h^\gamma_\gamma) - \nabla_\mu
\nabla^\beta h_{\beta \alpha} ] \,,
\end{eqnarray}
which is sourced by
\begin{eqnarray}
\bar{T}^{\;\; (1)}_{\mu\alpha} &=& {c \over 2 (q-1)!} f_{\mu \beta_2
\cdots \beta_q} \epsilon_\alpha^{\;\;\; \beta_2 \cdots \beta_q} - {c^2
(q-1) \over 2 (D-2)}\, h_{\mu \alpha} \\ &=& {c \over 2} \nabla_\mu
\nabla_\alpha b + {c \over 2} ( \square_y b_{\mu\alpha}
-  R_\alpha^{\;\; \beta} b_{\mu\beta}) - {c^2 (q-1) \over 2 (D-2)}\, h_{\mu \alpha} \,.
\end{eqnarray}
For $R_{\alpha\beta}$ we have
\begin{eqnarray}
R_{\alpha\beta}^{\;\; (1)} = -{1\over 2} [ (\square_x + \square_y) h_{(\alpha \beta)} - 2 R_{\alpha \gamma \delta \beta} h^{(\gamma \delta)} - R_\alpha^{\;\; \gamma} h_{(\gamma \beta)} - R_\beta^{\;\; \gamma} h_{(\gamma \alpha)} \\
+ {1 \over q} g_{\alpha \beta} (\square_x + \square_y) h^\gamma_\gamma
- ({2 \over q} + {2 \over p-2}) \nabla_\alpha \nabla_\beta h^\gamma_\gamma 
+ \nabla_\alpha \nabla_\beta H^\mu_\mu - \nabla_\alpha \nabla^\mu h_{\mu \beta} - \nabla_\beta \nabla^\mu h_{\mu \alpha}] \nonumber \,,
\end{eqnarray}
while on the right-hand side, we find
\begin{eqnarray}
\nonumber
\bar{T}^{\;\; (1)}_{\alpha\beta} = {c \over 2 (q-1)!} (f_{\alpha
\gamma_2 \cdots \gamma_q} \epsilon_\beta^{\;\;\; \gamma_2 \cdots
\gamma_q} + f_{\beta \gamma_2 \cdots \gamma_q} \epsilon_\alpha^{\;\;\;
\gamma_2 \cdots \gamma_q}) + {c^2 (q-1)\over 2 (q-1)!} (- h^{\gamma
\delta}) \epsilon_{\alpha \gamma \theta_3 \cdots \theta_q}
\epsilon_{\beta \delta}^{\;\;\;\;\; \theta_3 \cdots \theta_q} \\- {c^2
(q-1) \over 2(D-2)}  (h_{(\alpha \beta)} + {1 \over q} g_{\alpha \beta}
h^\gamma_\gamma ) - {c (q-1) \over (D-2)} g_{\alpha \beta} (f \cdot
\epsilon) - {q(q-1) c^2 \over 2 (D-2) q!} g_{\alpha \beta} (-
h^{\gamma \delta}) \epsilon_{\gamma \theta_2 \cdots \theta_q}
\epsilon_\delta^{\;\;\; \theta_2 \cdots \theta_q} \\
= {p-1 \over D-2} g_{\alpha \beta} \square_y cb + {q-1 \over R^2} h_{(\alpha \beta)} - {(q-1)^2 \over q R^2} g_{\alpha \beta} h^\gamma_\gamma \,, \nonumber
\end{eqnarray}
where we have used $(f \cdot \epsilon) = \square_y b$ and
$f_{\alpha_1 \cdots \alpha_q} = (f \cdot \epsilon)
\epsilon_{\alpha_1 \cdots \alpha_q} = \epsilon_{\alpha_1 \cdots
\alpha_q} \square_y b$.

We see that the modes of the graviton mix with the form modes $b$ and
$b_\mu$.  To solve the coupled systems, we must consider certain form
equations as well.  From the $\nabla^M F_{M \beta_2 \ldots \beta_q}$
equation\footnote{One may avoid explicit manipulation of Christoffel
symbols by linearizing the equivalent equation $\partial_M \sqrt{-g}
F^{M N_2 \cdots N_q} = 0$.}, we find the expression
\begin{eqnarray}
\label{FirstFormSphere}
\nabla^M f_{M \beta_2 \cdots \beta_q} - c g^{\mu \nu}
\Gamma_{\mu\nu}^{\gamma \, (1)} \epsilon_{\gamma \beta_2 \cdots
\beta_q} - c g^{\gamma \delta} \Gamma_{\gamma \delta}^{\theta \, (1)}
\epsilon_{\theta \beta_2 \cdots \beta_q} - c (q-1) g^{\gamma \delta}
\Gamma_{\gamma \beta_2}^{\theta \, (1)} \epsilon_{\delta \theta
\beta_3 \cdots \beta_q} = 0 \,,
\end{eqnarray}
where we use the linearized Christoffel symbol,
\begin{eqnarray}
\Gamma_{MN}^{P \, (1)} = {1\over 2} \left( \nabla_M h^P_N + \nabla_N
h^P_M - \nabla^P h_{MN} \right) \,.
\end{eqnarray}
Contracting with the epsilon tensor on $M_q$, (\ref{FirstFormSphere}) becomes 
\begin{eqnarray}
\label{FirstFormSphere2}
(q-1)! \left( \nabla_\alpha [( \square_x + \square_y) b + {c \over 2} H^\mu_\mu
- {c (p-1) \over p-2} h^\gamma_\gamma] + \nabla^\mu [\square_y b_{\mu \alpha} - R_{\alpha}^{\; \beta} b_{\mu \beta} - c h_{\mu \alpha}] \right) = 0 \,.
\end{eqnarray}
Finally, from the $\nabla^M F_{M \mu \beta_3 \ldots \beta_q}$ equation,
\begin{eqnarray}
\nabla^M f_{M \mu \beta_3 \cdots \beta_q} -c g^{\gamma \alpha}
\Gamma_{\gamma \mu}^{\delta \, (1)} \epsilon_{\alpha \delta \beta_3
\cdots \beta_q}= 0 \,,
\end{eqnarray}
which reduces to
\begin{eqnarray}
\label{FirstFormMixed}
(q-2)! \left[ \left(\square_x + \square_y - {2 (q-1) \over R^2}
\right) \nabla_{[\alpha} b_{\beta]\mu} - \nabla^\nu \nabla_\mu
\nabla_{[\alpha} b_{\beta]\nu} - c \nabla_{[\alpha} B_{\beta] \mu} + 2
R_{\alpha \;\; \beta}^{\;\; \gamma \;\; \delta} \nabla_{[\gamma}
b_{\delta] \mu} \right] \\ - (q-2)! D_{\beta_3} D^\nu a_{\mu \nu
\beta_4 \cdots \beta_q} \epsilon_{\alpha \beta}^{\;\;\;\;\; \beta_3
\cdots \beta_q} + (q-2)! (\square_x \beta_\mu - \nabla^\nu \nabla_\mu
\beta_\nu) = 0 \,. \nonumber
\end{eqnarray}
We now expand these fields in spherical harmonics and collect like
terms.  Below we present the results, collecting related equations and
indicating the origin of each expression as follows: (E1), (E2) and
(E3) for the $AdS$, mixed and $M_q$ Einstein equations, and (F1) and
(F2) for the form equations (\ref{FirstFormSphere2}) and
(\ref{FirstFormMixed}), respectively.

Equations for the coupled scalars $\pi^I$, $b^I$ and $H^I$:
\begin{eqnarray}
\rm{(E3)}& \left[ \left( \square_x + \square_y - {2 (q-1)^2 \over R^2}
\right)\pi^I + \square_y \left( H^I - {2 (D-2) \over q (p-2)} \pi^I
\right) + {2 q (p-1) \over (D-2)} \square_y cb^I \right] Y^I = 0
\,, \label{einsteinspheretrace} \\
\rm{(E3)}& \left( H^I - {2 (D-2) \over q
(p-2)} \pi^I \right) \nabla_{(\alpha} \nabla_{\beta)} Y^I = 0 \,,
\label{constraint} \\ 
\rm{(F1})& \nabla_\alpha \left( \square_x b^I + \square_y b^I + {c
\over 2} H^I - {c (p-1) \over (p-2)} \pi^I \right) Y^I = 0 \,,
\label{formspherescalar}
\end{eqnarray}
Equations for coupled vectors $b^I_\mu$, $B^I_\mu$:
\begin{eqnarray}
\label{einsteinmixedvector}
\rm{(E2)}& \left( {\rm Max}\ B_\mu^I + \Delta_y B_\mu^I + \Delta_y c
b_\mu^I - { 2 (q-1)^2 \over (p-1) R^2} b_\mu^I \right) Y_\alpha^I =
0 \,, \\ \rm{(F2)}& \nabla_{[\alpha} \left( {\rm Max}\ b_\mu^I +
\Delta_y b^I_\mu - c B_\mu^I \right) Y_{\beta]}^I = 0
\,. \label{formmixed2} \\ \rm{(F1)}& \left( \nabla^\mu b^I_\mu
\Delta_y - c \nabla^\mu B_\mu^I \right) Y^I_\alpha = 0 \,,
\label{formspherevector}\\ \rm{(E3)}& (\nabla^\mu B_\mu^I)
\nabla_{(\alpha} Y_{\beta)}^I = 0\,, \label{divergence}
\end{eqnarray}
Equations for symmetric tensors $H^I_{\mu \nu}$:
\begin{eqnarray}
\label{einsteinads}
\rm{(E1)}& (R_{\mu\nu}^{\;\; (1)} (H^I_{\rho \sigma}) - {1 \over 2} \square_y
H^I_{\mu \nu}  + {(q-1)^2 \over (p-1) R^2} H^I_{\mu \nu} +\\ & {1 \over 2
(p-2)} g_{\mu \nu} (\square_x + \square_y) \pi^I - {(q-1)^2
\over (p-2) R^2} g_{\mu \nu} \pi^I + { (q-1) \over
(D-2)} g_{\mu \nu} \square_y c b^I) Y^I = 0 \,, \nonumber \\
\rm{(E2)}&  \left(-
\nabla^\nu H_{\nu \mu}^I + \nabla_\mu H^I - { (p + q - 2) \over q
(p-2)} \nabla_\mu \pi^I + \nabla_\mu c b^I \right) \nabla_\alpha Y^I =0
\,, \label{einsteinmixedscalar} 
\end{eqnarray}
Note that in (\ref{einsteinads}), $R_{\mu\nu}^{\;\; (1)}$ is the
linearized Ricci tensor for $AdS_p$ only, evaluated on the field
$H_{\rho \sigma}$.  Finally, there remain a few decoupled equations:
\begin{eqnarray}
\rm{(E3)}& \left[ (\square_x + \square_y)
\delta_\alpha^\gamma \delta_\beta^\delta - 2 R_{\alpha \;\;\;\;
\beta}^{\;\; \gamma \delta} \right] \phi^I Y^I_{(\gamma \delta)} = 0
\,. \label{einsteinspheretraceless} \\
\rm{(F2)}& ({\rm Max}\ \beta_\mu^h ) Y^h_{[\alpha \beta]} = 0 \,, 
\label{harmonicvectors}\\
\rm{(F2)}& (\nabla^\nu b^I_{\nu \mu} ) \nabla_{[\alpha}
\nabla_{\beta]} Y^I = 0\,. \label{coclosed}
\end{eqnarray}
Notice that in passing from (\ref{FirstFormMixed}) to
(\ref{formmixed2}), we commuted the $\square_y$ through the covariant
derivative $\nabla_\alpha$, which not only produced precisely the
Laplacian $\Delta_y$ acting on vectors, but also canceled all terms in
(\ref{FirstFormMixed}) involving the Riemann tensor.  

It is worth remarking that as a result, the properties of $M_q$ enter
into almost all these formulas only through the dimension $q$ and the
radius $R$.  Consequently we will be able to treat these equations in
a completely unified way, and prove that for generic $AdS_p \times
M_q$ backgrounds, all the corresponding modes satisfy the
Breitenlohner-Freedman bound and cannot destabilize the background.
The sole exception is the equation (\ref{einsteinspheretraceless}) for
the scalars coming from graviton modes on the compact space, which
explicitly involves the Riemann tensor on $M_q$.  There is thus no
guarantee that the modes $\phi^I$ will possess the uniform stability
properties for different choices of $M_q$.  Indeed, we will find that
for $M_q = S^q$ these modes are harmlessly positive mass for all $q$,
while for any product $M_q = M_n \times M_{q-n}$ with $q < 9$ they
contain an instability.

\setcounter{equation}{0}
\section{Coupled scalars}
\label{ScalarSec}

In this section, we consider the system of modes associated with the
coupled scalars $\pi^I$, $b^I$ and $H^I$, equations
(\ref{einsteinspheretrace}), (\ref{constraint}) and
(\ref{formspherescalar}).

For certain low-lying scalar spherical harmonics $Y^I$, some or all of
their derivatives appearing in the equations of section
\ref{EquationSec} may vanish.  Let us first treat the generic case
where all derivatives of $Y^I$ in (\ref{einsteinspheretrace}),
(\ref{constraint}) and (\ref{formspherescalar}) are nonzero and hence
the coefficients must vanish.  Equation (\ref{constraint}) then gives us a
constraint which may be used to eliminate $H^I$ in favor of $\pi^I$.
Substituting into equation (\ref{formspherescalar}), we find
\begin{eqnarray}
\label{usedconstraint}
\left((\square_x + \square_y)\, b^I - c {(q-1) \over q} \pi^I \right)
Y^I = 0 \,,
\end{eqnarray}
while the second term in parentheses vanishes in (\ref{einsteinspheretrace}).
We obtain from
(\ref{einsteinspheretrace}) and (\ref{usedconstraint}) the coupled system
 \eqn{MassMatrix}{\eqalign{
 L^2 \square_x \pmatrix{ b^I/c \cr \pi^I} = 
  (p-1)^2 \pmatrix{ {\lambda^I \over (q-1)^2} & {R^2 \over q(q-1)} \cr
   {4 q \lambda^I \over (q-1) R^2} & {\lambda^I \over (q-1)^2} + 2}
  \pmatrix{ b^I/c \cr \pi^I } \,,
 }}
where $\square_y Y^I = - \lambda^I Y^I / R^2$; that $\lambda^I \geq 0$ is straightforward and is shown in the appendix.
On diagonalizing this matrix we obtain the mass spectrum 
 \eqn{Spectrum}{
 m^2 L^2 = {{(p-1)^2} \over {(q-1)^2}}[\lambda + (q-1)(q-1 \pm \sqrt{4 \lambda  + (q-1)^2})] \,.
 }
We now wish to analyze the spectrum (\ref{Spectrum}) to check
stability.  Extrema of (\ref{Spectrum}) occur for
\begin{eqnarray}
\label{Extrema}
1 \pm 2 (q-1) (4 \lambda + (q-1)^2)^{-1/2} = 0\,,
\end{eqnarray}
To satisfy (\ref{Extrema}) we must choose the negative sign, and we
find a minimum at
\begin{eqnarray}
\label{minimum}
\lambda = {3 \over 4} (q-1)^2 \,.
\end{eqnarray}
Substituting into (\ref{Spectrum}), we find the elegant result that
the minimum mass of the negative branch exactly saturates the
Breitenlohner-Freedman bound independent of $p$ and $q$,
\begin{eqnarray}
m^2_{min} L^2 = - \frac14 (p-1)^2 = m^2_{BF} L^2 \,.
\end{eqnarray}
Since the positive branch leads to manifestly positive masses, we have
proven there can be no unstable modes in this sector, at least for
modes associated to generic spherical harmonics.  We shall complete
the proof by treating the special cases momentarily.

Although the spectrum (\ref{Spectrum}) always saturates the BF bound
as a smooth function of $\lambda$, there need not be physical states at
the minimum, since only discrete values of $\lambda$ appear for given
$M_q$.  If $M_q = S^q$, then the eigenvalues of the spherical
harmonics are $\lambda = k(k+q-1)$, for integer $k \geq 0$, and the
mass formulas for the two branches take on the form \eqn{Mass}{ m_-^2
L^2 = {(p-1)^2 \over (q-1)^2} k (k-q+1) \,, \quad \quad m^2_+ L^2 =
{(p-1)^2 \over (q-1)^2} (k + 2(q-1))(k+q-1) \,.  } The minimum
(\ref{minimum}) occurs for $S^q$ at $k = (q-1)/2$ in the minus branch.
We notice that whenever $q$ is odd, there will be a mode with
precisely the Breitenlohner-Freedman mass, while for $q$ even the
lightest-mass states from this sector will appear just above the
bound.  This is consistent with what is already known about $AdS_4
\times S^7$ and $AdS_7 \times S^4$ \cite{Biranetal,Castellanietal,vanN}.

Let us now examine the special cases.  For $k=1$ on $S^q$,
$\nabla_{(\alpha} \nabla_{\beta)} Y^I = 0$ and we cannot use
(\ref{constraint}); this only occurs for maximally symmetric spaces,
and hence is not a concern for other $M_q$, where nonconstant $Y^I$
can be treated as above.  Following \cite{vanN} we may deal with this in
one of two ways: either using a residual gauge invariance to impose
the constraint anyway, or explicitly evaluating the remaining
equations and showing that one linear combination drops out.  We shall
do the latter; for a discussion of the former, see \cite{Kimetal}.

We now consider equation (\ref{formspherescalar}) as a constraint to
eliminate $H^I$ in favor of $\pi^I$ and $b^I$.  The remaining equation
(\ref{einsteinspheretrace}) becomes
\begin{eqnarray}
\left(\square_x + {3q-2 \over q} \square_y - {2 (q-1)^2 \over R^2}
\right) \pi^I - {R^2 (p-1) \over (q-1) (D-2)} \left( \square_x +
\square_y - {2q(q-1) \over R^2} \right) \square_y cb^I = 0 \,.
\end{eqnarray}
In the case of the sphere, $\square_y = - q/R^2$ and we find an
equation for a single mode,
\begin{eqnarray}
\left( \square_x - {q (2q-1) \over R^2} \right) \left( \pi^I + {q (p-1) \over (q-1)(D-2)} c b^I \right) = 0 \,,
\end{eqnarray}
which has the same mass as one would obtain from naively substituting
$k=1$ into the positive branch of (\ref{Mass}).  

For constant $Y^I$ on any $M_q$, all derivatives of $Y^I$ vanish and
the only nontrivial equation is (\ref{einsteinspheretrace}), which
reduces to
\begin{eqnarray}
\label{zerokscalar}
\left(\square_x - {2 (q-1)^2\over R^2} \right) \pi^I = 0 \,,
\end{eqnarray}
where again the mass matches what one obtains by substituting
$k=\lambda=0$ into the positive branch of (\ref{Spectrum}).  Thus we
learn that a proper treatment extends the positive branch of
(\ref{Spectrum}) down to $k=0$, while the negative branch truncates at
$k=2$ for $S^q$ and $k=1$ for other $M_q$.

The only remaining scalar fields associated to modes of the graviton
are the $\phi^I$, which obey the uncoupled equation
(\ref{einsteinspheretraceless}).  These shall turn out to be the modes
that can threaten stability.  We shall return to these in
section~\ref{UnstableSec}.

\setcounter{equation}{0}
\section{Coupled vectors}
\label{CoupledVectors}

We now consider the graviphoton $B_\mu$ and the form mode $b_\mu$ with
which it mixes.  We expect to find a massless vector for each Killing
vector on $M_q$ as well as a tower of massive fields, and indeed this
is what we obtain.  An additional $b^2(M_q)$ massless vectors
arise from the gauge potential, where $b^2(M_q)$ is the
second Betti number.

The relevant equations are (\ref{einsteinmixedvector}),
(\ref{formmixed2}), (\ref{formspherevector}), and (\ref{divergence}).
One readily sees that (\ref{formspherevector}) can be obtained from
the divergence of (\ref{formmixed2}).  We obtain the following coupled
system from (\ref{einsteinmixedvector}) and (\ref{formmixed2}):
\begin{eqnarray}
L^2\ {\rm Max} \pmatrix{ cb^I_\mu \cr B_\mu^I} = (p-1)^2 \pmatrix{
  {\kappa^I \over (q-1)^2} & {2 (D-2) \over (p-1)(q-1)} \cr {
  \kappa^I \over (q-1)^2} & {\kappa^I  \over (q-1)^2} + {2 \over p-1}}
  \pmatrix{ c b^I_\mu \cr B_\mu^I } \,, \label{VectorMass}
\end{eqnarray}
where $\Delta_y Y^I_{\alpha} = - \kappa^I Y^I_\alpha/R^2$.  The masses
that result are
\begin{eqnarray}
\label{vectormass}
m^2 L^2 = {(p-1)^2 \over (q-1)^2} \kappa + (p-1) \left( 1 \pm \sqrt{1 +
2 {p-1 \over (q-1)^3} (D-2) \kappa } \right) \,.
\end{eqnarray}
On a general Einstein space, we may derive the bound $\kappa^I \geq 2
(q-1)$, with equality when $Y^I_\alpha$ is a Killing vector, by
considering $\int d^qy \, S_{\alpha \beta} S^{\alpha \beta} \geq 0$
with $S_{\alpha \beta} \equiv \nabla_\alpha Y_\beta + \nabla_\beta
Y_\alpha$ (see appendix). For these Killing modes, the masses on the negative
branch of (\ref{vectormass}) vanish.  Hence we do indeed find a
massless vector for each isometry of the compact space $M_q$.  For
Killing modes (\ref{divergence}) is trivially satisfied and does not
constrain the vector fields.

The positive branch for $\kappa = 2 (q-1)$ yields a positive mass, and
one can show that for both branches (\ref{vectormass}) monotonically
increases with $\kappa$ for $\kappa \geq 2(q-1)$.  Thus all vector
modes are either massless or have positive mass.  For the non-Killing
modes (\ref{formspherevector}) and (\ref{divergence}) provide the
usual divergence-free condition for massive vectors, while for the
massive modes associated to the Killing vectors
(\ref{formspherevector}) accomplishes this by itself.

When the cohomology $H^2(M_q)$ is nontrivial, harmonic 2-forms
$Y^h_{[\alpha \beta]}$ give rise to $b^2(M_q)$ additional massless vectors
$\beta^h_\mu$, as we see from equation (\ref{harmonicvectors}).

\setcounter{equation}{0}
\section{Graviton and tensor fields}
\label{Graviton}

We now establish the existence of the $p$-dimensional graviton and
demonstrate the stability of the tower of massive symmetric two-index
tensors.  The graviton comes from the constant $Y^I$ mode of equation
(\ref{einsteinads}).  Using (\ref{zerokscalar}), this reduces to
\begin{eqnarray}
\label{AdSgraviton}
R_{\mu\nu}^{\;\; (1)} (H^I_{\rho \sigma}) + {p-1 \over L^2} H^I_{\mu
\nu} = 0\,,
\end{eqnarray}
which is the correct fluctuation equation for a linearized graviton in
$AdS_p$.

For generic $Y^I$, the trace and longitudinal parts of
(\ref{einsteinads}) are redundant given (\ref{einsteinspheretrace}),
(\ref{constraint}) (\ref{formspherescalar}), and
(\ref{einsteinmixedscalar}), which express the trace and divergence of
$H_{\mu\nu}$ in terms of $\pi$ and $b$.  One can use these equations
to reduce (\ref{einsteinads}) to
\begin{eqnarray}
\label{TensorEqn}
\left[ (\square_x + \square_y + {2 \over L^2}) H^I_{(\mu\nu)} - 2
\nabla_{(\mu} \nabla_{\nu)} c b^I \right] Y^I = 0\,.
\end{eqnarray}
A massive tensor field of mass $m^2$ is described by a field $\phi_{(\mu\nu)}$
which satisfies the wave equation and transversality constraints
 \begin{eqnarray}
(\square_x - m^2) \varphi_{(\mu\nu)} = 0 \,, &   \label{waveMT} \\
\nabla^\mu \varphi_{(\mu\nu)} = 0 \,. &          \label{transverseMT}
\end{eqnarray}
 To bring (\ref{TensorEqn}) to this form, we follow \cite{vanN}.  Define
$\phi_{(\mu\nu)}$ in terms of $H_{(\mu\nu)}$ by
\begin{equation}
H_{(\mu\nu)} = \phi_{(\mu\nu)} + \nabla_{(\mu} \nabla_{\nu)}(ub +v
\pi) \,,
\label{hformMT}
\end{equation}                    
where $u$ and $v$ are constants which can be determined by the
following procedure, which we outline without full detail. The first
step is to substitute \eno{hformMT} into (\ref{TensorEqn}) and require that
$\phi_{(\mu\nu)}$ satisfy \eno{waveMT} with mass $m^2_I =
\lambda^I/R^2 -2/L^2$, where $-\lambda^I/R^2$ is as usual the
eigenvalue of $\square_y$ on $Y^I$. The remaining terms are required
to cancel which gives one condition to determine $u$ and $v$. The
second condition is obtained by applying $\nabla^{\mu}$ to
\eno{hformMT}. The left side is expressed in terms of $b$ and $\pi$
using (\ref{constraint}) and (\ref{einsteinmixedscalar}), and one
imposes \eno{transverseMT}. After commuting covariant derivatives, one
finds two scalar conditions. Both contain the term
$\square_x(ub+v\pi)$ which may be eliminated between them. The
constants $u$ and $v$ may then be obtained by requiring that
coefficients of the independent fields $b(x)$ and $\pi(x)$ vanish. The
results are
\begin{eqnarray} u &= {2c(D-2)(p-2) \over
(q-1)L^2({\lambda^I \over R^2} - {p-2 \over L^2})} \\ v &=-{D-2 \over
q(p-1) ({\lambda^I \over R^2}- {p-2 \over L^2})}.  \end{eqnarray}
Strictly speaking the argument above does not apply to the $k=1$
graviton mode on $S^q$ since it uses the constraint (\ref{constraint})
which no longer follows from the Einstein equations. The simplest way
to extend the argument is to use the unfixed conformal diffeomorphisms
discussed in \cite{vanN} to impose the constraint for $k=1$. The
argument then goes through unchanged.

The apparent tensor mass $m_I^2$ is not positive for all geometries
$AdS_p \times M_q$. However \cite{vanN} one can examine $R_{\mu
\nu}^{\;\; (1)}$ in (\ref{AdSgraviton}) to see that the 
graviton itself has an apparent mass $-2/L^2$. When this is subtracted
one sees that higher tensor modes have positive mass
$\lambda^I/R^2$. These modes transform in unitary representations of
the $AdS_p$ isometry group, and we have stability.

\setcounter{equation}{0}
\section{Uncoupled form fluctuations}
\label{UncoupledSec}

As we saw, the gauge potentials with zero and one indices on $AdS_p$
mix with the graviton scalars and vectors.  The remaining form fields
are decoupled.  It is easiest to treat them using a differential form
notation.  Thanks to the gauge condition (\ref{CompactGauge}), these
may be written
\begin{eqnarray}
a(x,y) = \sum_I b^I(x) *_q d_q Y^I(y) + \sum_h \beta^h(x) Y^h(y) \,.
\end{eqnarray}
The linearized equation of motion is simply
\begin{eqnarray}
\label{LinearizedForm}
d * d a = 0\,.
\end{eqnarray}
Consider first the form $Y^I(y)$ with $n \geq 2$ indices on $M_q$; the
field $b^I$ then has $n$ indices on $AdS_p$.  Evaluating
(\ref{LinearizedForm}) and using the identities $ * (A_m(x) B_n(y) ) =
(-1)^{n(p-m)} *_p (A_m) *_q (B_n)$ and $d_q *_q Y^I = 0$, we arrive at
the equations
\begin{eqnarray}
\label{genericform}
(d_p *_p d_p b^I) d_q Y^I + (-1)^{n^2} (*_p b^I) d_q \Delta_y Y^I
= 0 \,, \\ (d_p *_p b^I) \Delta_y Y^I = 0 \,. \label{bcoclosed}
\end{eqnarray}
Equation (\ref{bcoclosed}) already appeared for the form with 2
indices on $AdS_p$ as (\ref{coclosed}).  It follows from
(\ref{bcoclosed}) that (\ref{genericform}) reduces to
\begin{eqnarray}
\left(\Delta_x - {\kappa^I \over R^2} \right) \, b^I=0  \,,
\end{eqnarray}
where $\Delta_x$ is the Laplacian on $AdS_p$ and $\Delta_y Y^I = -
\kappa^I Y^I / R^2$ is the eigenvalue of the Laplacian on $M_q$, as
before.  Thus these are standard positive-mass modes resulting from
the dimensional reduction.

The harmonic modes are even simpler; we find
\begin{eqnarray}
(d_p *_p \beta^h) Y^h = 0\,.
\end{eqnarray}
Thus we have a massless form of appropriate rank for each cohomology
class, as expected.

One potential modification of the action (\ref{Action}) is the addition
of a Chern-Simons term
\begin{eqnarray}
\label{CSAction}
\Delta S \sim \int A_{q-1} \wedge (F_q)^n \,,
\end{eqnarray}
where the wedge product is understood.  Naturally, this is only
possible when $q$ is even, and when an integer $n$ satisfying $nq = p
+1$ can be found.  (For $p=23$, $q=4$, one may add a CS term with
$n=6$.)  Notice that such a term breaks the duality symmetry between a
theory with $F_q$, which we have used, and a dual $F_p$; results for
the rest of this paper would be identical had we used $F_p$, but not
in this instance.  The modified action (\ref{CSAction}) leaves
Einstein's equations unchanged, and modifies the form equation to
\begin{eqnarray}
\label{CSEqn}
d * F_q = \gamma (F_q)^n \,,
\end{eqnarray}
for some constant $\gamma$.  In supersymmetric theories like
11-dimensional supergravity, the constant $\gamma$ is fixed by
supersymmetry.  Absent supersymmetry or some other guiding principle,
there is no preferred choice of $\gamma$.  For $n \geq 2$ our solution
(\ref{ProdSpace}), (\ref{BackgroundF}) is still valid since $F_q
\wedge F_q$ vanishes.  (For $n=1$, on the other hand, the Freund-Rubin
background is not a solution.)  Because $F_q \wedge F_q$ vanishes,
(\ref{CSEqn}) will begin to differ from (\ref{FormEOM}) only at the
$n-1$ order in perturbations.  Hence, our linearized analysis will
only be affected if $n=2$. Furthermore, for $f_q \wedge F_q$ to be
nonzero, the fluctuation $f_q$ must be polarized entirely along
$AdS_p$.  Hence, the addition of the term (\ref{CSAction}) can affect
our analysis for only the single mode (\ref{AllAdSb}).  We find the
equation
\begin{eqnarray}
\label{FinalCSEqn}
(\Delta_x + \Delta_y - 2 c \gamma *_p d_p) \,  b^I \, Y^I = 0\,.
\end{eqnarray}
We notice that $(*_p d_p)^2 = \Delta_x$ (for dimensions where
(\ref{CSAction}) is possible).  We can thus factorize (\ref{FinalCSEqn}) into
\begin{eqnarray}
\label{FactorCS}
(*_p d_p + m_1) (*_p d_p + m_2) \, b^I  Y^I = 0 \,, \\
m_1 + m_2 = - 2 c \gamma \,, \quad \quad m_1 m_2 =  - \kappa/R^2 \,,
\nonumber
\end{eqnarray}
with the solution
\begin{eqnarray}
m_1 = - c \gamma + \sqrt{c^2 \gamma^2 +
{\kappa \over R^2}} \,, \qquad
m_2 = - c \gamma - \sqrt{c^2 \gamma^2 +
{\kappa \over R^2}} \,.
\end{eqnarray}
There will be two towers, one annihilated by each of the factors in
(\ref{FactorCS}).  The second-order equations are
\begin{eqnarray}
(\Delta_x - m_i^2) \, b^I \, Y^I = 0 \,,
\end{eqnarray}
for $i=1,2$, and we see that $m_i^2$ are non-tachyonic masses
regardless of $\gamma$.

\setcounter{equation}{0}
\section{Metric perturbations on $M_q$ and stability}
\label{UnstableSec}

All the modes we have considered thus far have masses within the
bounds for stability; moreover, we were able to show this for $AdS_p
\times M_q$ where $M_q$ is an arbitrary $q$-dimensional Einstein
manifold.  The only fields we have not considered as yet come from the
traceless modes of the graviton on $M_q$, and satisfy
(\ref{einsteinspheretraceless}), which we repeat here:
\begin{eqnarray}
 \left[ (\square_x + \square_y) \delta_\alpha^\gamma
\delta_\beta^\delta - 2 R_{\alpha \;\;\;\; \beta}^{\;\; \gamma \delta}
\right] \phi^I\ Y^I_{(\gamma \delta)} &=& 0 \,. \label{phieqn}
\end{eqnarray}
It is possible to rewrite equation (\ref{phieqn}) in terms of the
Lichnerowicz operator $\Delta_L$ and the Ricci tensor:
\begin{eqnarray}
[ \square_x + \Delta_L + {2 (q-1) \over R^2} ] \phi^I\ Y^I_{(\alpha
\beta)} &=& 0 \,. \label{Lphieqn}
\end{eqnarray}
but since $\Delta_L$ does not obey a universal inequality as
$\square_y$ and $\Delta_y$ do, this form is not as useful.  The
presence of the Riemann tensor indicates that (\ref{phieqn}) can have
different properties depending on the particular choice of $M_q$.  We
give two examples, the sphere $M_q = S^q$ and a product space $M_q =
M_n \times M_{q-n}$, and show that the former contains only stable
modes while the latter possesses an instability for $q < 9$.

For $M_q = S^q$, the Riemann tensor has the maximally symmetric form
$R_{\alpha \beta \gamma \delta} = (g_{\alpha \gamma} g_{\beta \delta}
- g_{\alpha \delta} g_{\beta \gamma}) / R^2$.  Equation (\ref{phieqn})
reduces to
\begin{eqnarray}
 \left[ (\square_x + \square_y) - {2 \over R^2} \right] \phi^I\
Y^I_{(\alpha \beta)} &=& 0 \,. \label{phieqnsphere}
\end{eqnarray}
All these modes are manifestly positive-mass.  We thus complete our
demonstration of the stability of the $AdS_p \times S^q$ background
for all $p$ and $q$.

In \cite{DNP} and \cite{BerkoozRey} it was pointed out that $AdS_4
\times M_n \times M_{7-n}$ and $AdS_7 \times S^2\times S^2$,
respectively, were unstable to a perturbation in which one compact
space becomes uniformly larger and the other smaller keeping the total
volume fixed.  We now generalize this to an arbitrary product of
Einstein spaces $M_q = M_n \times M_{q-n}$ with $n \geq 2$.  Let $a,b$
denote indices on $M_n$ and $i,j$ denote indices on $M_{q-n}$.  If the
radii of the spaces are ${R}_1$ and ${R}_2$, requiring that the total
compact space is also Einstein imposes the relation
\begin{eqnarray}
\label{BackgroundProduct}
{n-1 \over {R}_1^2} = {q-n-1 \over {R}_2^2} = {q-1 \over R^2} \,.
\end{eqnarray}
Consider now the mode
\begin{eqnarray}
\label{unstablefluct}
h_{ab} = {1 \over n} g_{ab} \phi(x) \,, \quad \quad h_{ij} = - {1
\over q-n} g_{ij} \phi(x) \,,
\end{eqnarray}
which satisfies $h^\alpha_\alpha= 0$ as well as the gauge condition
(\ref{DDGauge}) and therefore obeys (\ref{phieqn}).  This perturbation
increases the radius of one of the Einstein spaces and decreases the
radius of the other keeping the total volume constant (to first
order).  Evaluating (\ref{phieqn}), we find
\begin{eqnarray}
\left[ \square_x + {2 (q-1) \over R^2} \right] \phi^I =  0 \,. 
\end{eqnarray}
Thus this mode has the mass
\begin{eqnarray}
m^2 L^2 = - {2 (p-1)^2 \over (q-1)} = {8 \over q-1} m^2_{BF} L^2 \,. 
\end{eqnarray}
Consequently the Breitenlohner-Freedman bound (\ref{Bound}) is violated
for $ q < 9$.  This result is independent of $p$, and depends on the
internal space only in that it is a product of Einstein spaces that is
itself Einstein with total dimension $q$; in particular the relative
dimension of the spaces is irrelevant.

One may wonder about other fluctuations obeying (\ref{phieqn}), and
whether they may place even more stringent constraints on the
requirements for stability.  The field (\ref{unstablefluct}) is the
lowest in a tower of modes that are traces on each individual space in
the product, but traceless overall.  Higher excitations will have more
positive masses from the $\square_y$ term.  The remaining modes are
traceless on each $M_n$ and $M_{q-n}$, namely $h_{(ab)}$, $h_{(ij)}$
and $h_{ai}$.  For $h_{ai}$ we find the universal result
\begin{eqnarray}
(\square_x + \square_y) h_{ai} = 0 \,,
\end{eqnarray}
which is obviously stable, while for either of the other two we have
effectively a copy of equation (\ref{phieqn}) but involving the
Riemann tensor of just one of the spaces in the product:
\begin{eqnarray}
 \left[ (\square_x + \square_y) \delta_a^c
\delta_b^d - 2 R_{a \;\;\;\; b}^{\;\; cd}
\right] h_{(cd)} &=& 0 \,. \label{phieqnrecursive}
\end{eqnarray}
and similar for $h_{(ij)}$.  This obviously depends on the details of
$M_n$.  One observation we can make is that if $M_n$ itself is a
product (and so the original compact space $M_q$ is a product of three
or more manifolds), a mode analogous to (\ref{unstablefluct}) will have
a mass $m^2 = 2(n-1)/{R}_1^2 = 2(q-1)/R^2$, where the last equality
comes from (\ref{BackgroundProduct}), and thus will be unstable
precisely when (\ref{unstablefluct}) is; and hence no new instability
automatically arises for products of three or more spaces beyond that
already generically present for a product of two.

\setcounter{equation}{0}
\section{$AdS_4$ vacua of massive IIA}
\label{MassiveIIA}

Massive type IIA supergravity has $AdS_4 \times M_6$ vacua
\cite{RomansIIA} which are non-supersymmetric and whose stability, to
our knowledge, has never been investigated.\footnote{There is also a
{\it supersymmetric} (and necessarily stable) vacuum which is a
fibration of $AdS_6$ over $S^4$ with a non-trivial dilaton.  It is the
near-horizon geometry of the D4-D8 system \cite{BrandhuberOz}.  It
would be interesting to explore the properties of this background as
well as generalizations of it where $S^4$ is replaced by other
manifolds, but we will not do so here.}  Even the existence of these
solutions is non-trivial, since there is a potential term for the
dilaton which pushes it toward weak coupling.  What makes $AdS_4
\times M_6$ vacua possible is that a uniform RR field strength, $F_4$
or $F_6$ according to taste, pushes the dilaton toward strong
coupling, and there is an extremum of this total potential where the
dilaton can be constant.

The extremum is in fact a maximum, but it doesn't make sense to ask
whether second derivative of the total dilaton potential alone
satisfies the BF bound, because the dilaton couples non-trivially to
the form and to the graviton.  This mixing means that the coupled
scalars sector requires a more intricate analysis than before.  The
result will be that the apparent $s$-wave tachyon coming from a naive
analysis of the dilaton potential is completely erased (effectively,
it is a gauge artifact), but for $S^6$ there is a $d$-wave and an
$f$-wave mode which violates the BF bound, rendering this vacuum
unstable! To our knowledge, this is the first time that a product of
AdS and a round sphere is unstable. We also show that for $M_6 = S^n
\times S^{6-n}$ the BF bound is violated within the coupled scalar
sector, as well as having the same purely gravitational instability
found earlier, where one factor shrinks while the other grows.

The remaining modes, outside the coupled scalar sector, satisfy the
same equations as in the generic $AdS_p \times M_q$ systems we already
considered.  Thus the traceless graviton on $M_6$ joins the coupled
scalars as a possible source of instability.  We do not analyze other
Einstein manifolds $M_6$ explicitly, but we provide the tools needed
for such an analysis.  It is still possible that there exist stable
$AdS_4 \times M_6$ vacua.

To make the discussion similar to our previous analysis,
let us express the action for massive IIA in terms of a six-form field
strength, which is essentially the Hodge dual of the usual four-form:
  \eqn{IIAaction}{
   S = {1 \over 2\kappa^2} \int d^{10} x \sqrt{g} \left[ R - 
    {1 \over 2} (\partial\phi)^2 - {1 \over 2} \xi^2 F_6^2 - 
    {m^2 \over 8} \xi^{-10} \right]  \qquad 
     \hbox{where $\xi = e^{-\phi/4}$} \,,
  }
 and we include a $1/6!$ in the definition of $F_6^2$, as in
\cite{PolchVolTwo}.  We also
include a factor of $1/q!$ in the inner product of forms, $\omega_q
\cdot \tilde\omega_q$.  The equations of motion are
  \eqn{IIAeoms}{\eqalign{
   &R_{MN} = {m^2 \over 64} \xi^{-10} g_{MN} + 
    {1 \over 2} \partial_M \phi \partial_N \phi + 
    {\xi^2 \over 2 \cdot 5!} F_{MP_1P_2P_3P_4P_5} F_N{}^{P_1P_2P_3P_4P_5} - 
    {5 \over 16} \xi^2 g_{MN} F_6^2 \,, \cr
   &\square\phi - {5 \over 16} m^2 \xi^{-10} + 
    {\xi^2 \over 4} F_6^2 = 0  \,, \cr
   & d * \xi^2 F_6 = 0 \,,
  }}
 and there is an $AdS_4 \times M_6$ background with $\phi=0$, $F_6 = c
\vol_{M_6}$.  We readily derive the relations
  \eqn{IIArelations}{
   c^2 = F_6^2 = {5 \over 4} m^2 = {10 \over L^2} = {25 \over R^2} \,,
  }
 where $L$ is the radius of curvature of $AdS_4$, such that\
$R_{\mu\nu} = -{3 \over L^2} g_{\mu\nu}$, and $R$ is the radius of
curvature of $M_6$, such that $R_{\alpha\beta} = {5 \over R^2}
g_{\alpha\beta}$.

Just as for $AdS_p \times M_q$, we wish to linearize around the
background to obtain the mass spectrum.  For the coupled scalar
sector, we wish to focus on perturbations of the form $g_{MN} \to
g_{MN} + h_{MN}$ with $h_{\mu\nu} = {1 \over 4} g_{\mu\nu}
h^\lambda_\lambda$ and $h_{\alpha\beta} = {1 \over 6} g_{\alpha\beta}
h^\gamma_\gamma$.  Also let $\delta\phi$ be the perturbation in $\phi$
and let $f_6$ be the perturbation in $F_6$, where, as before, we write
\eqn{alphaDef}{\eqalign{ &h^\alpha_\alpha = \pi \,, \quad f_6 = da_5
\,, \quad {\rm where} \quad a_5 = *_6 db \,.  }} The algebraic
relation $h^\mu_\mu + h^\alpha_\alpha = {1 \over 3} h^\alpha_\alpha$
follows from the symmetric traceless part of the Einstein equations,
as before.  It is now possible to derive coupled second order
equations relating $\delta\phi$, $b$, and $\pi$ from the variations of
the $R^\alpha_\alpha$ Einstein equation, the scalar equation, and the
form equation, using the algebraic relation when needed to eliminate
$h^\mu_\mu$ in favor of $h^\alpha_\alpha$.  We use a form notation
in this section for convenience.

The $R^\alpha_\alpha$ equation is
  \eqn{Reom}{
   R^\alpha_\alpha = {3 \over 4L^2} \xi^{-10} + {9 \over 8} \xi^2 F_6^2
    + {1 \over 2} \partial^\alpha \phi \partial_\alpha \phi \,.
  }
 Using \RicciExpand, \IIArelations, and \alphaDef, we find
  \eqn{varR}{\eqalign{
   \delta R^\alpha_\alpha &= -{2 \over L^2} h^\alpha_\alpha - 
    {1 \over 2} (\square_x + \square_y) h^\alpha_\alpha - 
    {1 \over 2} \square_y (h^\mu_\mu + h^\alpha_\alpha) + 
    {1 \over 6} \square_y h^\alpha_\alpha  \cr
    &= -{15 \over 4 L^2} \delta\phi + {9 \over 4} c \square_y b - 
     {45 \over 4 L^2} h^\alpha_\alpha \,,
  }}
 where we have used the fact that $\square_y = *_6 d *_6 d$ acting on
$b$.  The algebraic relation allows us to simplify this to
  \eqn{finalR}{
   (\square_x + \square_y) \pi - {37 \over 2L^2} \pi - 
    {15 \over 2 L^2} \delta\phi + {9 \over 2} c \square_y b = 0 \,.
  }
For the scalar, the equation of motion is
  \eqn{scalareom}{
   \square \phi - {5 \over 2L^2} \xi^{-10} + 
    {\xi^2 \over 4} F_6^2 = 0 \,.
  }
 Linear variation around the background gives
  \eqn{varscalar}{
   (\square_x + \square_y) \delta\phi - {25 \over 4L^2} \delta\phi
    - {1 \over 8} \delta\phi F_6^2 + {1 \over 2} F_6 \cdot f_6 - 
    {1 \over 4} h^{\alpha\beta} F_{\alpha\gamma_1\ldots\gamma_5} 
     F_\beta{}^{\gamma_1\ldots\gamma_5}
     {1 \over 5!} = 0 \,,
  }
 which upon simplification and use of $F_6 \cdot f_6 = c \square_y b$
becomes
  \eqn{finalscalar}{
   (\square_x + \square_y) \delta\phi - {15 \over 2L^2} \delta\phi
    + {1 \over 2} c \square_y b - {5 \over 2L^2} \pi = 0 \,.
  }
The variation of the form equation is
  \eqn{Formeom}{
   d (\delta *) F_6 - d * {1 \over 2} \delta\phi F_6 + d * f_6 = 0 \,,
  }
 where $\delta *$ indicates the variation in the Hodge dual.  After
some algebra this becomes
  \eqn{varForm}{
   {c \over 2} d (h^\mu_\mu - h^\alpha_\alpha - \delta\phi) 
    \wedge \vol_4 + 
   d (\square_x + \square_y) b \wedge \vol_4 = 0 \,,
  }
 and so, using the algebraic relation, we obtain
  \eqn{finalForm}{
   (\square_x + \square_y) b - {5c \over 6} \pi - 
    {c \over 2} \delta\phi = 0 \,.
  }

Gathering everything together, setting $b = cL^2B$ for convenience,
and recalling that $c^2 = 10/L^2$, one obtains the
following system of equations:
  \eqn{ThreeSystem}{\eqalign{
   & (\square_x + \square_y) B - {5 \over 6 L^2} \pi - 
    {1 \over 2 L^2} \delta\phi = 0  \cr
   & (\square_x + \square_y) \pi - {37 \over 2 L^2} \pi - 
     {15 \over 2L^2} \delta\phi + 45 \square_y B = 0  \cr
   & (\square_x + \square_y) \delta\phi - {5 \over 2 L^2} \pi - 
    {15 \over 2 L^2} \delta\phi + 5 \square_y B = 0 \,.
  }}
 This results in 
 \eqn{IIAMassMatrix}{\eqalign{
   L^2 \square_x \pmatrix{ B \cr \pi \cr \delta\phi} = 
   \pmatrix{ {2 \over 5} \lambda & {5 \over 6} & {1 \over 2} 
      \cr\noalign{\vskip1\jot}
    {18 \lambda} & {2 \over 5} \lambda + {37 \over 2} & {15 \over 2} 
     \cr\noalign{\vskip1\jot}
    2 \lambda & {5 \over 2} & {2 \over 5} \lambda + {15 \over 2}}
   \pmatrix{ B \cr \pi \cr \delta\phi} \,,
  }}
 where as before $-R^2 \square_y Y^I =\lambda Y^I$.  We find the mass
eigenvalues
  \eqn{IIAmasses}{
   m^2 L^2 = {2 \over 5} \lambda + 6 \,, \quad
     {2 \over 5} \lambda + 10 + 2 \sqrt{25 + 4 \lambda} \,, \quad \hbox{and} \ 
     {2 \over 5} \lambda + 10 - 2 \sqrt{25 + 4 \lambda} \,.
  }
The Breitenlohner-Freedman bound for $p=4$ is $m^2 L^2 \geq -9/4$.  We
see that the first two towers in \IIAmasses\ are harmless (in fact
they're not even tachyonic), but the third tower will violate the BF
bound if some value of $\lambda$ falls in the interval
  \begin{eqnarray}
   \label{UnstableInterval}
    \lambda_{\rm unstable} \in \left( {155 \over 8} - 5\sqrt{5 \over 2}, 
     {155 \over 8} + 5\sqrt{5 \over 2} \right) \approx
      (11.47,27.28) \,.
  \end{eqnarray}
 For $S^6$, we have $\lambda = k(k+5)$, for which $k=2,3$ gives values
in the interval (\ref{UnstableInterval}).  Thus for both $d$- and
$f$-waves, the eigen-combinations of $B, \pi,$ and $\delta \phi$
corresponding to the third eigenvalue in (\ref{IIAmasses}) are
unstable modes of the $AdS_4 \times S^6$ solution. They have the
common mass $m^2 L^2=-12/5$.

It is interesting that in fact all values of $m^2 L^2$ that occur for
$AdS_4 \times S^6$ in the coupled scalar sector are rational: upon
substituting $\lambda = k(k+5)$ into \IIAmasses, we obtain
  \eqn{IIAmagain}{
   m^2 L^2 = {2k^2 \over 5} + 2k + 6 \,, \quad
    {2k^2 \over 5} + 6k + 20 \,, \quad \hbox{and} \ 
    {2k^2 \over 5} - 2k \,.
  }
 However the corresponding dimensions of operators in a hypothetical
three-dimensional CFT are not rational.

Instabilities can occur in the coupled scalar sector of other $M_q$ as
well.  As an example, consider $M_6 = S^n \times S^{6-n}$.  For
product spherical harmonics on the two spheres labeled by $(k_1,
k_2)$, we find several unstable modes in the interval
(\ref{UnstableInterval}): $(1,1)$, $(0,2)$ and $(1,2)$ for $n=2$, and
$(1,1)$, $(2,0)$ and $(0,2)$ for $n=3$.

As in section \ref{ScalarSec}, the constraint relating $h^\mu_\mu$ and
$h^\alpha_\alpha$ no longer obtains for the $k=1$ case on $S^6$, so a
more careful analysis must be performed.  Without imposing the
algebraic constraint, the dilaton equation (\ref{finalscalar}) is
unmodified, while equations (\ref{varR}) and (\ref{varForm}) become
\begin{eqnarray} (\square_x + \square_y) \pi + \square_y (H + \pi) -
{1 \over 3} \square_y \pi - {37 \over 2L^2} \pi - {15 \over 2 L^2}
\delta\phi + {9 \over 2} c \square_y b = 0 \,, \label{REqnGen}\\
(\square_x + \square_y) b - {c \over 2} \pi + {c \over 2} H - {c \over
2} \delta\phi = 0 \,. \label{FormEqnGen}
\end{eqnarray}
For $k=1$, we have $\square_y = - 12/5 L^2$.  The dilaton equation
(\ref{finalscalar}) then becomes
\begin{eqnarray}
\label{FinalEqnOne}
(\square_x - {99 \over 10 L^2} ) \delta \phi - {5 \over 2 L^2} \pi -
{6 \over 5 L^2} cb &=& \nonumber \\ (\square_x - {99 \over 10 L^2}
) \delta \phi - {5 \over 2 L^2} \sigma &=& 0 \,,
\end{eqnarray}
which defines $\sigma \equiv \pi + {12 \over 25} cb$.  Next, using (\ref{FormEqnGen}) we can show that
\begin{eqnarray}
\square_x \pi + \square_y H = \square_x \sigma - {12 \over 5 L^2} \sigma
- {12 \over 5 L^2} \delta \phi \,,
\end{eqnarray}
which allows us to write equation (\ref{REqnGen}) as
\begin{eqnarray}
\square_x \sigma - {249 \over 10 L^2} \sigma - {99 \over 10 L^2}
\delta \phi  = 0\,.
\label{FinalEqnTwo}
\end{eqnarray}
As in the examples without a coupled dilaton, one linear combination
of fields has dropped out of the $k=1$ system.  We can now diagonalize
the equations (\ref{FinalEqnOne}) and (\ref{FinalEqnTwo}).  We
discover the mass eigenvalues
\begin{eqnarray}
m^2 L^2 = {42 \over 5} \,, \quad \quad m^2 L^2 = {132 \over 5} \,,
\label{OneMasses}
\end{eqnarray}
 which coincide with the $k=1$ masses in the first two towers of
\IIAmagain.

The constant $Y^I$ sector is straightforward for all $M_6$.  The form
equation no longer obtains, and the $b$ mode does not exist, leaving
only the equations
\begin{eqnarray}
\square_x \pi = {37 \over 2 L^2} \pi + {15 \over 2 L^2} \delta \phi
\,, \\ \square_x \delta \phi = {15 \over 2 L^2} \delta \phi + {5 \over
2 L^2} \pi \,,
\end{eqnarray}
with corresponding positive-mass eigenvalues
\begin{eqnarray}
m^2 L^2 = 6 \,, \quad \quad m^2 L^2 = 20 \,.
\end{eqnarray}
 These are exactly the masses obtained from the first two towers in
(\ref{IIAmasses}) with $k=\lambda=0$.  Thus the general ``rule of
thumb'' (valid in all cases we have considered, as well as in the
familiar supersymmetric examples) is that one simply drops the most
tachyonic mode from the first two partial waves in the coupled scalar
sector.

It is not hard to see that the remaining equations of motion are
basically unmodified from the analysis of previous sections.  The
dilaton fluctuation $\delta \phi$ cannot appear in the other
polarizations of the form equation, where the background field
strength vanishes.  Hence these are unchanged from before.  In the
Einstein equations, it is straightforward that $\delta \phi$ does not
appear in the $R_{\mu \alpha}$ equation or in parts of the $R_{\alpha
\beta}$ equation other than those treated already by considering the
trace.  Owing to the relations (\ref{IIArelations}) arising from the
requirement that the compact space is Einstein, these equations are
identical to those we already studied once written in terms of $L$.
The dilaton fluctuation and the other scalars do appear in the
$R_{\mu\nu}$ equation, analogous to the appearance of $\pi$, $b$ and
$H$ in (\ref{einsteinads}), but this leads only to a scalar expression
linearly dependent on the ones we have considered earlier.

Consequently, we can employ the work we have already done wholesale.
In particular, we again have the potential source of instability from
the set of scalars $\phi^I$, obeying equation (\ref{phieqn}).  Hence
we learn that general product spaces are again unstable against having
one factor shrink while the other grows.

\setcounter{equation}{0}
\section{Possible CFT duals}
\label{CFTDuals}

As discussed in the introduction, this investigation was motivated by
the proposal \cite{HS} that the case $D=p+q=27$ with a 4-form field is
the low-energy limit of a ``bosonic M-theory," and that its $AdS_4
\times S^{23}$ compactification has a CFT$_3$ dual in the framework of
the $AdS$/CFT correspondence. Since an $AdS_p \times S^q$
compactification has been shown to be stable, it is interesting to
speculate in general about possible CFT$_d$ duals (with $d=p-1$).  We
give a very heuristic discussion which emphasizes the pattern of
operator dimensions.

For scalar operators the basic $AdS$/CFT relation
$\Delta(\Delta-d)=m^2 L^2$ admits the two roots
\begin{equation}
\Delta_{\pm} = {d \over 2} \pm {1 \over 2} \sqrt{d^2 + 4m^2L^2} \,.
\end{equation}
 If the mass satisfies the inequality $m^2L^2 \geq -{d^2 \over 4} +
1$, then only the assignment $\Delta_+$ obeys the unitarity bound
$\Delta \geq {d \over 2}-1$.  (This bound is saturated for a free
massless scalar field in $d$ dimensions).  But for $-{d^2 \over 4}
\leq m^2L^2 \leq -{d^2 \over 4} + 1$, both $\Delta_+$ and $\Delta_-$
are, {\it a priori}, consistent choices for the scale dimension of the
dual operator.  On general grounds it seems most natural to choose the
larger of the two dimensions, $\Delta_+$, as the dimension of the
operator, because only then can one compute correlators by
straightforwardly imposing a boundary condition on the larger of the
two linearly independent solutions of the scalar.  If $\Delta_-$ is
chosen as the dimension, then to obtain field theory correlators one
must make a Legendre transform of the $\Delta_+$ results.  These
points were discussed in \cite{KlebWit}, where also a particular
example was exhibited where the $\Delta_-$ dimension was needed.  In
this example, the field theory was supersymmetric, and the operator
was a chiral primary, so its anomalous dimension could be worked out
purely on field theory grounds as the sum of the the anomalous
dimensions of its factors.  The computation is rigorous because all
the dimensions are dictated by a $U(1)_R$ current which is obviously
additive.

The mass eigenvalues of coupled scalars of general $AdS_p \times S^q$
compactifications are given in (\ref{Mass}). Since $m^2_+ >0$, the
operator duals of positive branch scalars have the unique dimension
assignments
\begin{equation}
\Delta ={ p-1 \over 2} \left[1 + {2 \over q-1} \left(k + {3(q-1) \over 2} \right) \right] \,.
\end{equation}
For the negative branch of the scalar mass spectrum, there are the
two possibilities
\begin{equation}
\label{operatordimminus}
\Delta_{\pm} ={ p-1 \over 2} \left[1 \pm {2 \over q-1} \left|k-{q-1
\over 2} \right| \right] \,.
\end{equation}
In accord with the discussion in the previous paragraph the negative
root is a possible choice in the range
\begin{equation}
\left|k -{q-1 \over 2} \right| \leq {q-1 \over d} \,.
\end{equation}
Recall that $k$ indicates the $SO(q+1)$ representation formed from $k$
factors of the vector, then symmetrized with the trace removed.

For the purposes of orientation, let us recall a familiar result for
$AdS_5 \times S^5$.  Here the chiral primary operators are $\tr
X^{(I_1} \cdots X^{I_k)}$ in ${\cal N}=4$ super-Yang-Mills theory,
where $(I_1 \ldots I_k)$ indicates the symmetric traceless
combination.  Their $AdS$ duals are the coupled fluctuations of the
metric and the five-form on the negative branch that leads to
\eno{operatordimminus}.  The dimensions are $\Delta(k) = k =
2,3,4,5,\ldots$, and one always chooses $\Delta_+$.  The anomalous
dimensions vanish: $\Delta(k) = k$ is the free-field result.  A
similar story holds for $AdS_4 \times S^7$, with $\Delta(k) = k/2$,
except that one must choose $\Delta_-$ for $k=2$.  Some of these
operators are thought of as coming from $\tr X^{(I_1} \cdots X^{I_k)}$
on coincident D2-branes, and for the others one must dualize the
vector boson into an eighth scalar.  Free field counting still
applies, and it can be backed up by a supersymmetry argument as for
the $AdS_5 \times S^5$ case.  Lastly, for $AdS_7 \times S^4$, the
dimensions are $\Delta(k) = 2k$, and one always chooses $\Delta_+$.  A
free field understanding is lacking in this mysterious $(2,0)$ theory,
but as before a link can be established between the R-symmetry and the
dimension which guarantees that $\Delta(k)$ is linear in $k$.

Let us begin the discussion of the spectra for general $p$ and $q$ by
observing that it is doubly remarkable that both the quadratic
equation for scalar masses and the equation $\Delta(\Delta-d)=m^2 L^2$
have rational roots in the general case.  This is an aesthetically
pleasing point for a putative CFT dual, but unfortunately it is the
end of the good news.

Focusing on the negative branch \eno{operatordimminus} makes sense,
since these were the simplest operators in cases which we understand.
Starting with our free field prejudices, we might suspect that the
$k$'th operator would be expressible as $\tr X^{(I_1} \cdots
X^{I_k)}$, and that its dimension $\Delta(k)$ is linear in $k$.  Then
we arrive at $\Delta(k) = {p-1 \over q-1} k$.  For example, $\Delta(k)
= {3 \over 22} k$ for $AdS_4 \times S^{23}$.  This does not make sense
because $k=2$ gives $\Delta = {3 \over 11} < {1 \over 2}$, the free
scalar dimension.  That is, we tried to choose $\Delta_-$ in a range
where only $\Delta_+$ was possible.  The general result is that a
linear spectrum of dimensions $\Delta(k)$ is permitted provided
  \eqn{linearcriterion}{
   q-1 \leq {4(p-1) \over p-3} \,.
  }
 If this inequality fails, as in the case $AdS_4 \times S^{23}$, then
some operators of low $SO(q+1)$ charge will have a larger dimension
than operators of higher $SO(q+1)$ charge, which we may view as a
failure of the free-field intuition that singlet operators are built
from fundamental fields whose dimensions add.  It does not mean,
however, that there can't be a CFT dual: for instance, it is
consistent with the unitarity bound to choose $\Delta_+$ uniformly,
which produces a spectrum $\Delta(k)$ with a kink about $k={q-1 \over
2}$.  More arcane choices may also be imagined.  In the absence of
supersymmetry or some input from field theory, we have no way of
deciding between the alternatives.

Let us now discuss the spectra of coupled vectors for general $AdS_p
\times S^q$ compactifications.  Inserting the eigenvalue formula
$\kappa = (k+1)(k+q-2)$ for vector spherical harmonics in
(\ref{vectormass}), we find the masses
 \begin{equation}
m^2 L^2 = {(p-1)^2 \over (q-1)^2} (k+1)(k+q-2) + (p-1) \left( 1 \pm \sqrt{1 +
2 {p-1 \over (q-1)^3} (p+q-2) (k+1)(k+q-2) } \right) \,.
\end{equation}
These mass eigenvalues are generically irrational (although they are
rational for the supersymmetric compactifications $AdS_4 \times S^7$,
$AdS_7 \times S^4$ and $AdS_5 \times S^5$.) Irrationality persists for
vector scale dimensions (except for Killing vectors, where $ m^2=0$)
\begin{equation} \Delta = {1 \over 2} [ d + \sqrt{(d-2)^2 + 4m^2}] \,.
\end{equation}
 In particular, $AdS_4 \times S^{23}$ has irrational masses and
dimensions for massive vectors.

It is certainly remarkable that the scalars dual to chiral primary
operators in the well-understood $AdS_5 \times S^5$, $AdS_4 \times
S^7$, and $AdS_7 \times S^4$ vacua still lead to rational dimensions
for general $p$ and $q$.  If \linearcriterion\ is violated and a
linear spectrum of dimensions is impossible for scalars, then it seems
difficult to imagine a concise understanding based on a Lagrangian.
The fact that massive vector modes generically have irrational
dimensions also makes it seem less likely that a purely field
theoretic formulation of the putative dual CFT will be accessible in
the near future.

The $AdS_4 \times S^6$ compactification presents an even less rosy
picture, in that the BF bound is violated.  Obvious candidates for a
brane realization of this vacua (involving D2-branes and D8-branes)
seem also to be unstable, only the instability is usually in the form
of a tadpole instead of a tachyon.  It would be very interesting if a
stable $AdS_4 \times M_6$ vacuum could be found for appropriate $M_6$,
corresponding to some analyzable type I$'$ brane configuration.  It
would also be satisfying if one could start with some unstable D2-D8
construction and show that in an appropriate near-horizon limit the
brane instability reduces to the violations of the BF bound that we
have observed.\footnote{We thank O.~Bergman and A.~Brandhuber for
discussions on these and related points.}

Finally, let us extend some remarks on thermodynamics made in
\cite{HS} for the $AdS_4 \times S^{23}$ and $AdS_{23} \times S^4$
cases.  An obvious measure of the number of degrees of freedom in a
CFT in $p-1$ dimensions is the ratio $c_{\rm thermo} = S/(VT^{p-1})$.
In the $p+q$-dimensional theory, there are solutions with both
magnetic and electric charge under the field strength $F_q$, so there
is flux quantization, and we can ask how $c_{\rm thermo}$ scales with
$N$, the number of flux quanta through the compact space.  For $AdS_p
\times S^q$, we can reason out this scaling by recalling that in an
asymptotically flat solution, the number of branes enters the harmonic
function in the metric as $H = 1 + c_1 N (\ell_{\rm Pl} / r)^{q-1}$,
where $c_1$ is some dimensionless constant.  Thus $L$ and $R$
scale as $N^{1/(q-1)} \ell_{\rm Pl}$.  In a near-extremal solution,
the Bekenstein-Hawking entropy scales as $(L/\ell_{\rm Pl})^{p+q-2}$,
whereas the Hawking temperature does not scale with $\ell_{\rm Pl}$ at
all.  Putting everything together, one finds
  \eqn{GotCThermo}{
   c_{\rm thermo} \sim N^{(p+q-2)/(q-1)} \,.
  }
 This specializes to the odd results $c_{\rm thermo} \sim N^{25/22}$
for $AdS_4 \times S^{23}$ and $c_{\rm thermo} \sim N^{25/3}$ for
$AdS_{23} \times S^4$.  These peculiar fractions do not bring any
known CFT's to mind, but at least they represent something to shoot
for in constructing putative duals of $AdS_p \times M_q$.

\setcounter{equation}{0}
\section{Implications for Extremal Black Branes and Negative Energy}
\label{Endpoint}

As mentioned above, the (nondilatonic) theories of gravity (\ref{Action})
all contain
charged black brane solutions, where the charge is obtained by
integrating $F_q$ over an $S^q$ surrounding the brane. (For the
general solution, see \cite{GHT}.) In particular, there are extremal
black branes, with metric
 \eqn{extbrane}
{ds^2 = H^{-{2\over p+1}} (-dt^2 + d{\bf y}\cdot d{\bf y})
+ H^{{2\over q-1}} (dr^2 + r^2
d\Omega_q)}
where $H$ is the harmonic function $H(r) = 1 + c_1 N(\ell_{\rm Pl}/r)^{q-1}$.
 The
near horizon limit is just $AdS_p
\times S^q$. So the stability we have found for $AdS_p \times S^q$ for all
$p$ and $q$ is consistent
with the expected stability of extremal solutions. However, we have also
seen that $AdS_p \times M_n \times M_{q-n}$ is unstable, when $q<9$ and
$M_n, \  M_{q-n}$ are Einstein spaces. These can also arise as the near
horizon limit of a type of extremal black brane as follows.
Consider the cone over $M_n \times M_{q-n}$
\eqn{cone}
{ds^2 = dr^2 + r^2(d\sigma_{M_n}^2 + d\sigma_{M_{q-n}}^2)}
This space is Ricci flat, and has a curvature singularity at the apex $r=0$.
(Even though the curvature goes to zero for large $r$, this space
 is not asymptotically flat in the usual sense since the curvature only
falls off like $r^{-2}$.) Suppose one places a stack of branes at
the apex of the cone, extended in the orthogonal directions. The resulting
exact solution is obtained by
simply replacing the flat transverse metric in (\ref{extbrane})
with the cone metric (\ref{cone}).

One might
have expected this new solution to be stable, since it is the extremal
limit of a family of black brane solutions.  However it is easy to see that
it is not (at least for $q<9$). The near horizon
limit is now $AdS_p \times M_n \times M_{q-n}$ which
is unstable to a perturbation (equation (\ref{unstablefluct}))
that goes to zero asymptotically in $AdS_p$. So a similar perturbation
with support very close to the horizon of the extremal black brane will
also grow exponentially. This is independent of the change in boundary
conditions at infinity since, in the Poincare coordinates appropriate to the
near horizon geometry of $AdS_p$, a scalar field near the horizon has
a unique evolution inside a spacetime region that includes
infinite Poincare time. One might object that extremal black branes
are always unstable in the sense that adding a small amount of energy
causes them to become nonextremal\footnote{This is true for branes of finite
extent. For infinite branes, one needs nonzero energy density to become
nonextremal.}, and the horizon moves from an infinite
distance to a finite distance (in spacelike directions).
 However, as we will see, our perturbation is
very different in that it can
actually decrease the mass.

A natural question to ask is what does this instability lead to? As we have
seen, the unstable mode causes one factor, say $M_n$, to shrink in size and
the other to grow. So one might expect that in the full nonlinear evolution,
$M_n$ simply shrinks to zero size.
However this cannot happen. It has recently been shown \cite{GH} that if
the weak energy condition is satisfied, event horizons cannot have
collapsing cycles. 
In fact, given any spacelike curve on the event horizon,
if one evolves the curve along the null geodesic generators, its length cannot
go to zero in finite affine parameter.
The basic idea is to use the fact that
the divergence $\theta$ of the null geodesic
generators $\ell$
of the event horizon cannot become negative. This means that if part of the
horizon
is contracting, the orthogonal directions must be expanding. But this
introduces a lot of shear $\sigma_{MN}$ in the null geodesic congruence.
One now uses the Raychaudhuri equation
\eqn{Rayd}{
{d \theta \over d\lambda} = -{\theta^2\over D-2}  -\sigma_{MN}\sigma^{MN}
- R_{MN} \ell^M \ell^N}
where  $\lambda$ is an affine parameter along the null geodesics and
$D$ is the total spacetime dimension. If the weak energy condition is
satisfied, the right hand side is negative definite, so when the shear
becomes large, $\theta$ decreases rapidly.
One can show
that if part of the horizon shrinks to zero size in finite affine parameter,
  $\theta$ must become negative
which is a contradiction. So the solution must settle down to a new
static configuration.
In \cite{GH}, this result was discussed in the context of the Gregory-Laflamme
instability of  {\it nonextremal} black branes.
In that case, the horizon starts to shrink in
some places and expand in others, and it was widely believed that the horizon
would eventually pinch off and form separate black holes. However  this
cannot happen. Instead, the solution must settle down to a new static black
brane solution without translational symmetry along the brane.

The instability we are discussing can be viewed as an extremal analog of the
Gregory-Laflamme instability. Since our theory satisfies the weak energy
condition, and the result in \cite{GH} does not require that
the horizon is nonextremal, it can also be applied to our  case.
Thus, $M_n$ cannot shrink to zero size, and there must be
another static solution whose near horizon geometry is not
$AdS_p \times M_n \times M_{q-n}$.
\footnote{One might worry that there will be a
problem applying the result in \cite{GH} since the unstable
extremal black brane is not asymptotically flat in the usual sense.
However, even though null infinity is not
well defined, one can still define the event horizon as the boundary
of the past of a surface at large $r$, and the result will still apply.}

Strictly speaking, the near horizon limit of the black brane solution
includes only part of $AdS_p$ (the region covered by the Poincare coordinates).
Suppose we now consider the global solution $AdS_p \times M_n \times M_{q-n}$
and ask what happens if we perturb it in the unstable direction. As a first
step toward answering this question, we show that there are solutions in the
full nonlinear theory which are asymptotically $AdS_p \times M_n
\times M_{q-n}$ and have arbitrarily negative energy (where, as usual,
we measure energy relative to $AdS_p$).  Since
the perturbation violates the BF bound, it is clear we can lower the energy
slightly by turning on this mode. To show the energy can be arbitrarily 
negative,
it suffices to construct suitable initial data. Consider the spatial metric
\eqn{inidata}{
ds^2 = \bigg[{r^2\over L^2} +1 - {m(r)\over r^{p-3}}\bigg]^{-1} dr^2
   + r^2 d\Omega_{p-2}+
 e^{(q-n)\phi(r)}d\sigma_{M_n} + e^{-n\phi(r)} d\sigma_{M_{q-n}}  }
so $m=0, \phi=0$  corresponds to the metric on a static surface
 (in global coordinates) for $AdS_p \times M_n \times M_{q-n}$.
The total mass is proportional to $m(\infty)$.
Notice that the volume of the $q$-dimensional internal space is independent of
$\phi$. This is a nonlinear generalization of the perturbation we
considered in section \ref{UnstableSec}. We again set $F_q = c \ {\rm vol}_{M_q}$. If we set all time derivatives to zero, the only constraint
on this initial data is the Hamiltonian constraint of general relativity
which implies that the scalar curvature of (\ref{inidata})
must be $c^2/2$ where $c^2$ is given by (\ref{Valueofc}). 
 This yields a first order differential equation which
can be used to solve for
$m(r)$ in terms of $\phi(r)$.
 If we assume $\phi$ is everywhere small, this equation
becomes
\eqn{massequ}{
{m'\over r^{p-2}}
 \propto \bigg[{r^2\over L^2} +1 - {m(r)\over r^{p-3}}\bigg] (\phi')^2 -
{2(p-1)^2\over (q-1)L^2} \phi^2  }
The right hand side resembles the energy density of the linearized
unstable mode (\ref{unstablefluct}) except that the
$\phi'$ term involves the corrected spatial metric. Since the term
involving $m(r)$ on the right hand side only decreases the energy
density we can get an upper limit on the mass by dropping it.  One can
now explicitly find $\phi(r)$ so that $m(\infty)$ is arbitrarily
negative. For example, if $q<9 - (8/p)$, one can take $\phi = \phi_0
e^{-r/a}$. The total mass is negative for large $a$, and goes to minus infinity as
$a \rightarrow \infty$.

If we start with $AdS_p \times M_n
\times M_{q-n}$ and perturb it slightly, the energy will be only
slightly negative. As we have just seen, this is very far from the minimum
energy solution.
A priori, one might
expect $M_n$ to collapse down to zero size in finite time. This will
produce a curvature singularity.  It is unlikely that this singularity is
naked, since we don't expect cosmic censorship to be violated so easily in
the higher dimensional theory of gravity we are considering. It may form
a black hole, or in light of the horizon results,
 $M_n$ may not collapse down at
all. In the latter case,
since we are using reflecting boundary conditions at infinity
(appropriate for the AdS/CFT correspondence), the solution may not settle down
to any static configuration.  It would be interesting
to investigate this further. 

We have not considered the massive IIA theory in this section.  It
would also be interesting to investigate the implications of the
instability of $AdS_4\times S^6$ for negative energies and extremal
black branes in this theory.

\section*{Acknowledgments}

We would like to thank O.~Bergman, I.~Klebanov, and E.~Witten for
useful discussions.  We are particularly grateful to A.~Brandhuber for
help checking some of the equations in the massive IIA
compactifications.  The work of S.~S.~G.\ and I.~M.\ was supported in
part by the DOE under grant DE-FG03-92ER40701 and through an
Outstanding Junior Investigator Award. I.~M.\ was also supported under
grant ~DE-FG02-91ER40671. The work of O.~D.\ was supported by the NSF
under grant PHY-99-07949.  The work of G.~H.\ was supported by the NSF
under grant PHY-00-70895.  The work of D.~Z.~F.\ was supported by the
NSF under grant PHY-97-22072. I.~M.\ thanks the Institute for
Theoretical Physics, Santa Barbara and the Particle Theory Group at
Caltech for their hospitality during most of the phases of this
project.

\appendix
\setcounter{equation}{0}
\section{Appendix}

Here we collect conventions and a few properties of the differential
operators we employ.  We work in a metric of signature $(-++ \cdots
+)$ and define the Ricci tensor in terms of the Riemann tensor by
$R_{MN} \equiv R^P_{\;\; MPN}$.

The Hodge-de Rham Laplacian $\Delta_y = - (d^\dagger d + d d^\dagger)$
is negative-definite, but in the case of a compact Riemannian Einstein
space of positive curvature a more stringent bound can be derived for
the case of one-forms.  We use $- R^2 \Delta_y Y^I \equiv \kappa^I
Y^I$, and for the ordinary Laplacian $\square_y \equiv g^{\alpha
\beta} \nabla_\alpha \nabla_\beta$, $- R^2 \square_y Y^I \equiv
\lambda^I Y^I$.  
For scalar spherical harmonics $Y^I$, $\square_y = \Delta_y$, and a
vanishing eigenvalue always exists corresponding to $Y^I =$
const.\footnote{One can derive the bound $\lambda^I \geq q$ for
nonconstant $Y^I$ \cite{Duffetal}.}  For one-forms, we may consider
\begin{eqnarray}
0 &\leq& \int (\nabla^\alpha Y^{I\beta} + \nabla^\beta Y^{I\alpha})
(\nabla_\alpha Y_{\beta}^I + \nabla_\beta Y^I_\alpha) = 2 \int
\nabla^\alpha Y^{I\beta} (\nabla_\alpha Y_{\beta}^I + \nabla_\beta
Y^I_\alpha) \\ &=& - 2 \int Y^{I\beta} (\square_y + {q-1 \over R^2})
Y^I_\beta = - 2\int Y^{I\beta} (\Delta_y + {2(q-1) \over R^2})
Y^I_\beta \,, \nonumber
\end{eqnarray}
proving $\kappa^I \geq 2(q-1)$; furthermore, equality occurs for
$(\nabla_\alpha Y_{\beta}^I + \nabla_\beta Y^I_\alpha) = 0$, which is
the condition for $Y_{\beta}^I$ to be a Killing vector.  Additionally,
the absence of harmonic one-forms $Y^h_\alpha$ on a compact Einstein
space of positive curvature may be proved as follows.  Any harmonic
one-form must satisfy $\nabla^\alpha Y^h_\alpha = 0 =\nabla_\alpha
Y_{\beta}^h - \nabla_\beta Y^h_\alpha$, so
\begin{eqnarray}
0 = \int \nabla^\alpha Y^{h\beta} (\nabla_\alpha Y_{\beta}^h -
\nabla_\beta Y^h_\alpha) = \int \left(\nabla^\alpha Y^{h\beta}
\nabla_\alpha Y^h_\beta + {q-1 \over R^2} Y^{h\beta} Y^h_\beta\right)
\,,
\end{eqnarray}
which is impossible as the right-hand side is a sum of a nonnegative
and a positive quantity.

For the case of $S^q$, the eigenvalues $\lambda^I$ of the ordinary
Laplacian $\square_y$ for the various tensor harmonics are
\begin{center}
\begin{tabular}{c|c|c} 
Tensor harmonic & $\lambda^I$ & Range of $k$  \\ \hline
$Y^I$ & $k(k+q-1)$ & $k \geq 0$ \\
$Y^I_\alpha$ & $k(k+q-1) -1 $&  $k \geq 1$ \\
$Y^I_{[\alpha_1 \cdots \alpha_n]}$ & $k(k+q-1) -n $ & $k \geq 1$ \\
$Y^I_{(\alpha \beta)}$ & $k(k+q-1) -2 $ & $k \geq 2$ \\
\end{tabular}
\end{center}
while for the Hodge-de Rham Laplacian acting on vectors, we obtain
\begin{eqnarray}
\kappa^I = (k+1)(k+q-2) \,, \quad \quad k \geq 1 \,.
\end{eqnarray}

\begingroup\raggedright\endgroup

\end{document}